%% file: ms.tex
\documentclass{emulateapj}
\usepackage{colordvi}
\usepackage{epsfig}
\usepackage{graphicx}
\usepackage{amssymb}

\newcommand{\mic}{~$\mu$m}

\newcommand{\Mg}{Mg$_{UV}$}

\begin{document}

\title{SHARDS: an optical spectro-photometric survey of distant galaxies}

\author{Pablo G. P\'erez-Gonz\'alez\altaffilmark{1,2}, Antonio
  Cava\altaffilmark{1}, Guillermo Barro\altaffilmark{1,3}, V\'{\i}ctor
  Villar\altaffilmark{1}, Nicol\'as Cardiel \altaffilmark{1}, Ignacio
  Ferreras\altaffilmark{4}, Jos\'e Miguel
  Rodr\'{\i}guez-Espinosa\altaffilmark{5}, Almudena
  Alonso-Herrero\altaffilmark{6,7}, Marc Balcells\altaffilmark{5,8},
  Javier Cenarro\altaffilmark{9}, Jordi Cepa\altaffilmark{5},
  St\'ephane Charlot\altaffilmark{10}, Andrea Cimatti\altaffilmark{11},
  Christopher J. Conselice\altaffilmark{12}, Emmanuele Daddi\altaffilmark{13},
  Jennifer Donley\altaffilmark{14}, David Elbaz\altaffilmark{13},
  N\'estor Espino\altaffilmark{1}, Jes\'us Gallego\altaffilmark{1}, R.
  Gobat\altaffilmark{15}, Omaira
  Gonz\'alez-Mart\'{\i}n\altaffilmark{5}, Rafael
  Guzm\'an\altaffilmark{16}, Antonio
  Hern\'an-Caballero\altaffilmark{6}, Casiana
  Mu\~noz-Tu\~n\'on\altaffilmark{5}, Alvio Renzini\altaffilmark{17},
  Javier Rodr\'{\i}guez Zaur\'{\i}n\altaffilmark{5}, Laurence
  Tresse\altaffilmark{18}, Ignacio Trujillo\altaffilmark{5}, Jaime
  Zamorano\altaffilmark{1}}

\altaffiltext{1}{Departamento de Astrof\'{\i}sica, Facultad de CC.
  F\'{\i}sicas, Universidad Complutense de Madrid, E-28040 Madrid,
  Spain}

\altaffiltext{2}{Associate Astronomer at Steward Observatory, The
  University of Arizona}

\altaffiltext{3}{UCO/Lick Observatory, Department of Astronomy and
  Astrophysics, University of California, Santa Cruz, CA 95064, USA}

\altaffiltext{4}{Mullard Space Science Laboratory, University College
  London, Holmbury St Mary, Dorking, Surrey RH5 6NT}

\altaffiltext{5}{Instituto de Astrof\'{\i}sica de Canarias, 38200 La
  Laguna, Tenerife; Departamento de Astrof\'{\i}sica, Universidad de
  La Laguna, E-38205 La Laguna, Tenerife, Spain}

\altaffiltext{6}{Instituto de F\'{\i}sica de Cantabria,
  CSIC-Universidad de Cantabria, 39005 Santander, Spain}

\altaffiltext{7}{Augusto Gonz\'alez Linares Senior Research Fellow}

\altaffiltext{8}{Isaac Newton Group of Telescopes, Aptdo. 321, 38700
  Santa Cruz de La Palma, Spain}

\altaffiltext{9}{Centro de Estudios de F\'{\i}sica del Cosmos de
  Arag\'on, Plaza San Juan 1, Planta 2, 44001 Teruel, Spain}

\altaffiltext{10}{Institut d'Astrophysique de Paris, CNRS, Universit\'e
  Pierre \& Marie Curie, UMR 7095, 98bis bd Arago, 75014 Paris,
  France}

\altaffiltext{11}{Dipartimento di Astronomia, Universit\'a degli Studi di
  Bologna, I-40127 Bologna, Italy}

\altaffiltext{12}{School of Physics \& Astronomy, University of
  Nottingham, Nottingham NG7 2RD, UK}

\altaffiltext{13}{CEA, Laboratoire AIM, Irfu/SAp, F-91191
  Gif-sur-Yvette, France}

\altaffiltext{14}{Los Alamos National Laboratory, Los Alamos, NM, USA}

\altaffiltext{15}{Laboratoire AIM-Paris-Saclay,
  CEA/DSM-CNRS-Universit\'e Paris Diderot, Irfu/Service d'Astrophysique,
  CEA Saclay, Orme des Merisiers, 91191 Gif-sur-Yvette, France}

\altaffiltext{16}{Department of Astronomy, University of Florida, 211
  Bryant Space Science Center, Gainesville, FL 32611, USA}

\altaffiltext{17}{INAF - Osservatorio Astronomico di Padova, Vicolo
  dell'Osservatorio 5, I-35122 Padova, Italy}

\altaffiltext{18}{Aix Marseille Universit\'e, CNRS, LAM 
  (Laboratoire d'Astrophysique de Marseille) UMR 7326, 13388, Marseille, France}

\slugcomment{Last edited: \today}
\date{Submitted: \today}


\label{firstpage}
\begin{abstract}

  We present the Survey for High-z Absorption Red and Dead Sources
  (SHARDS), an ESO/GTC Large Program carried out with the OSIRIS
  instrument on the 10.4m Gran Telescopio Canarias (GTC). SHARDS is an
  ultra-deep optical spectro-photometric survey of the GOODS-N field
  covering 130~arcmin$^2$ at wavelengths between 500 and 950~nm with
  24 contiguous medium-band filters (providing a spectral resolution
  R$\sim$50). The data reach an AB magnitude of 26.5 (at least at a
  3$\sigma$ level) with sub-arcsec seeing in all bands.  SHARDS main
  goal is obtaining accurate physical properties of intermediate and
  high-z galaxies using well-sampled optical SEDs with sufficient
  spectral resolution to measure absorption and emission features,
  whose analysis will provide reliable stellar population and AGN
  parameters. Among the different populations of high-z galaxies,
  SHARDS principal targets are massive quiescent galaxies at z$>$1,
  whose existence is one of the major challenges of current
  hierarchical models of galaxy formation. In this paper, we outline
  the observational strategy and include a detailed discussion of the
  special reduction and calibration procedures which should be applied
  to the GTC/OSIRIS data. An assessment of the SHARDS data quality is
  also performed. We present science demonstration results about the
  detection and study of emission-line galaxies (star-forming objects
  and AGN) at z$=$0--5.  We also analyze the SEDs for a sample of 27
  quiescent massive galaxies with spectroscopic redshifts in the range
  1.0$<$z$\lesssim$1.4. We discuss on the improvements introduced by
  the SHARDS dataset in the analysis of their star formation history
  and stellar properties.  We discuss the systematics arising from the
  use of different stellar population libraries, typical in this kind
  of studies. Averaging the results from the different libraries, we
  find that the UV-to-MIR SEDs of the massive quiescent galaxies at
  z$=$1.0--1.4 are well described by an exponentially decaying star
  formation history with scale $\tau$$=$100--200~Myr, age around
  1.5-2.0~Gyr, solar or slightly sub-solar metallicity, and moderate
  extinction, A(V)$\sim$0.5~mag. We also find that galaxies with
  masses above M$^\ast$ are typically older than lighter galaxies, as
  expected in a downsizing scenario of galaxy formation. This trend
  is, however, model dependent, i.e., it is significantly more evident
  in the results obtained with some stellar population synthesis
  libraries and almost absent in others.

\end{abstract}

\section{Introduction}
\label{sect:intro}

The current paradigm of galaxy formation establishes that the baryons
closely follow the evolution of the Cold Dark Matter (CDM) halos,
which cluster and grow hierarchically as shown in cosmological
simulations and semi-analytical models (such as those in
\citealt{2005Natur.435..629S}; see also
\citealt{1998ApJ...498..504B,1999MNRAS.310.1087S,2000MNRAS.319..168C,
  2008MNRAS.391..481S,2010A&A...518A..14R}).  In this scenario, star
formation started within the cooling gas clouds in merging dark matter
halos after a relatively slow early collapse regulated by feedback
processes.  This early star formation produced relatively small disk
systems that later merged and generated larger (i.e., more massive)
spheroidal systems \citep[see,
e.g.,][]{1993MNRAS.264..201K,2000A&G....41b..10E,2001ApJ...560L.119E}

The global picture about the co-evolution of matter in the Universe
(including all graviting components: CDM and baryons) is
self-consistent and has been successful in reproducing and even
predicting many observables about galaxy evolution, especially at low
redshift. Among the most relevant successes, we find the good
comparison of models with the observed power spectrum of the Cosmic
Microwave Background \citep{2007ApJS..170..377S,2011ApJS..192...18K}
or the Large Scale Structure of the Universe
\citep{2001MNRAS.327.1297P,2001MNRAS.328.1039C}. Also very convincing
is the link between observations and theoretical expectations such as
the existence and properties of the acoustic baryonic oscillations
\citep{2005ApJ...633..560E,2009MNRAS.399.1663G,2010MNRAS.401.2148P}.
In addition, the hierarchical scenario for galaxy formation is also
supported by the observations of galaxy mergers at different
cosmological distances
\citep[e.g.,][]{1999ApJ...520L..95V,2005AJ....130.2647V,2006ApJ...650L..29M},
and the increase of the fraction of galaxies undergoing mergers as we
move to higher redshifts \citep[among
others,][]{1993MNRAS.262..627L,2000MNRAS.311..565L,2003AJ....126.1183C,
  2006ApJ...652..270B,2008ApJ...672..177L,2009A&A...501..505L}.

However, the hierarchical picture contrasts with several observational
evidences, especially at high redshift (z$>$1--2), where a more
classical monolithic collapse is favored. This formation path was
proposed 50 years ago to explain the origin of bulges such as the
Milky Way's and spheroidal galaxies. This would be done through a
free-fall rapid collapse causing the formation of the bulk of the
stars in these systems in a short period of time. Later, the star
formation is shut off by some {\it quenching} phenomena, and the
galaxy henceforth evolves passively
\citep{1962ApJ...136..748E,1974MNRAS.166..585L}.  This theory was
largely abandoned due, first, to the compilation of evidence
supporting that spheroidal galaxies suffer merging episodes
\citep{1972ApJ...178..623T}. Furthermore, globular clusters and the
general stellar population in the MW present a relatively wide range
of ages \citep[e.g.,][]{1978ApJ...225..357S}, directly pointing out to
the hierarchical scenario. Eventually, the hierarchical picture was
adopted instead of the monolithic collapse due to the high degree of
success of the $\Lambda$CDM models and semi-analytic models mentioned
above.

Nevertheless, rapid early episodes of intense star formation are
indeed consistent (although not uniquely) with observational facts in
nearby galaxies, such as the dominant old stellar populations in
bulges and ellipticals, their metallicity and $\alpha$-elements
enhancement, and the dynamics and shape of these systems \citep[e.g.,
][]{1997ApJS..111..203V,1997AJ....114.1771F,2000AJ....119.1645T,
  2000AJ....120..165T}. In addition, hierarchical models still present
severe drawbacks in several aspects. The most challenging
observational facts for hierarchical models refer to the lightest and
heaviest galaxies.  Indeed, hierarchical models typically present a
``missing satellite problem'', i.e., they predict many more low-mass
galaxies than what is actually observed
\citep[see][]{1993MNRAS.264..201K,1999MNRAS.303..188K,1999ApJ...522...82K,
  2004ApJ...609..482K,2009MNRAS.398.2177L,2012ApJ...752L..19Q}. On the
bright end, models tend also to overpredict the number of massive
galaxies observed in the local Universe, although they are getting
closer to the observations after taking into account quenching
mechanisms
\citep{2006MNRAS.365...11C,2007MNRAS.375....2D,2008MNRAS.391..481S,
  2010A&A...518A..14R,2010MNRAS.404.1111G}.

The discrepancies between the predictions of current galaxy formation
models based on the $\Lambda$CDM paradigm and the data are more
obvious as we move to higher redshifts. In the last 15 years, a wide
variety of papers using very heterogeneous data and methods have
presented compelling evidence that the formation of galaxies follows a
so-called {\it downsizing} scenario
\citep{1996AJ....112..839C,2004Natur.428..625H,2004Natur.430..181G,
  2005ApJ...621L..89B,2005ApJ...630...82P,2007A&A...476..137A,2008ApJ...675..234P}.
In this theory, the most massive galaxies formed first in the history
of the Universe, thus having the oldest stellar populations seen
today.  The formation of less massive systems continued at lower
redshifts.  Downsizing implies that the bulk of the star formation in
the most massive galaxies happened quick and stopped for some reason
in early times.  This also means that there should be massive
passively evolving galaxies at high redshift. This kind of objects
have indeed been detected at redshifts around z$\sim$1--3 with a
variety of techniques \citep{2000AJ....120..575Y2,2003ApJ...587L..79F,
  2004ApJ...617..746D,2006ApJ...640...92P,2008A&A...482...21C}.

The finding of massive galaxies at z$=$1--3, some of them already
evolving passively, is indeed extremely challenging for current models
of galaxy formation based on the $\Lambda$CDM paradigm. Indeed, models
predict many less massive systems at high-z than observed \citep[see,
e.g.,][]{2007MNRAS.381..962C,2009ApJ...701.1765M,2012MNRAS.421.2904H,
  2012ApJ...744..159L}.  In contrast, the downsizing scenario
contradicts, at least at first sight, the predictions of a
hierarchical assembly of the stellar mass in galaxies, i.e., the most
massive galaxies do not seem to be the result of multiple mergers
occurring in a extended period along the Hubble time
\citep{1998ApJ...498..504B,2000MNRAS.319..168C, 2007ApJ...665..265F}.
Still, a hierarchical assembly with (maybe multiple) mergers occurring
at high redshift between gas-rich systems (a process close in nature
to a monolithic collapse) would be consistent with both the evidences
for downsizing and the properties of the dominant stellar populations
seen in nearby spheroidal systems \citep[e.g.][]{2009Natur.457..451D}.

From the observational point of view, our understanding of the
processes involved in the early (z$>$1) assembly of galaxies (and also
the evolution from the early Universe to the present) is still
hampered by the significant (often systematic) uncertainties in our
estimations of their physical properties. Our global picture of galaxy
formation will only improve if we are able to get more robust
estimations of some key properties of galaxies, such as the stellar
masses, SFRs, and extinctions. Jointly with those, we of course need
better estimations of the distances to galaxies based on spectroscopic
or photometric redshifts, which can be used to relate the mentioned
galaxy properties with other relevant parameters such as the
environment. The improvements in the determination of stellar masses
and SFRs/extinctions will also mean a better estimation of the age of
the stellar population and the Star Formation Histories, SFH
\citep[see, e.g,][]{2001ApJ...559..620P,2006A&A...459..745F,
  2008ApJ...677..219K,2008A&A...477..503E,2012MNRAS.tmp.2944P,2012MNRAS.421.2002P}.
Jointly with this observational effort, models should also be
improved, including better physics. For example, models are still to
provide more certain emissivities of the stellar populations in the
rest-frame NIR, now affected by strong uncertainties due to
limitations of about knowledge about the properties and importance of
stellar evolutionary phases such as the thermally-pulsating TP-AGB
phase \citep[see][]{2005MNRAS.362..799M,2010ApJ...722L..64K}. The task
of obtaining more robust physical parameters of galaxies at
cosmological distances is even more interesting for those massive
galaxies which have already reached a quiescent state and are evolving
passively at high-z, whose number densities and properties are the
most demanding challenges for current galaxy evolution models. The
cosmological importance of these systems is very high, since they most
probably represent the early formation phases of present-day
early-type galaxies.

In this paper, we present the basics of the {\it Survey for High-z
  Absorption Red and Dead Sources} (SHARDS), an ESO/GTC Large Program
awarded 180 hours of GTC/OSIRIS time during 2010-2013.  This project
consists of an ultra-deep (m$<$26.5~AB mag) imaging survey in 24
medium-band filters covering the wavelength range between 500 and
950~nm and targeting the GOODS-N field. The observations carried out
by SHARDS allow to accurately determine the main properties of the
stellar populations present in these galaxies through
spectro-photometric data with a resolution R$\sim$50, sufficient to
measure absorption indices such as the D(4000)
\citep[e.g.,][]{1983ApJ...273..105B,1999ApJ...527...54B,
  2003MNRAS.341...33K,2011ApJ...743..168K} or \Mg\, index
\citep{1997ApJ...484..581S,2004ApJ...614L...9M,2005MNRAS.357L..40S,
  2005ApJ...626..680D,2008A&A...482...21C}. The analysis of these
spectral features is a powerful method to constrain the solutions of
stellar population synthesis models and to improve our estimations of
parameters such as the age, SFH, mass, and extinction of galaxies at
cosmological distances.

SHARDS inherits the observational strategy of past and on-going
optical surveys such as COMBO17
\citep{2001A&A...377..442W,2003A&A...408..499W}, the COSMOS
medium-band survey \citep{2009ApJ...690.1236I}, ALHAMBRA
\citep{2008AJ....136.1325M}, and PAU/J-PAS
\citep{2009ApJ...691..241B,2011arXiv1108.2657A}.  These projects have
demonstrated the impact of large photometric datasets on our
understanding of the formation of galaxies (see, among many papers,
\citealt{2004ApJ...608..752B,2007ApJ...665..265F,2004A&A...421..913W,
  2004ApJS..152..163R,2006A&A...453..869B,2006ApJ...637..727C,
  2007ApJS..172..150S,2009ApJ...692L...5B,2010ApJS..189..270C,2011ApJ...735...86W}).
SHARDS intends to be a step forward from these surveys in terms of
depth, spectral resolution, and data quality. Our survey prioritizes
the detailed study of the faintest galaxies at the highest redshifts
over the analysis of closer galaxy populations and the Large Scale
Structure at intermediate redshift, and thus focuses on a smaller area
than the surveys mentioned above.  Indeed, SHARDS was planned to reach
up to 3 mag fainter than those surveys, uses typically twice the
number of filters in the same wavelength range (i.e., our spectral
resolution is better), and was obtained in excellent (sub-arcsec)
seeing conditions with a 10m class telescope. In contrast, it covers a
fraction of the area surveyed by other projects.

In this paper, we present the main technical characteristics of the
survey in Section~\ref{sect:shards_description}, including a thorough
discussion of the reduction and calibration procedures in
Section~\ref{sect:shards_data}.  Next, we present our science
verification results about emission-line and absorption systems. In
Sections~\ref{sect:elgs} and \ref{sect:laes}, we discuss about the
ability of the SHARDS data to select and study emission-line sources
(star-forming galaxies and AGN) at intermediate (z$<$1) and high
redshifts (up to z$\sim$5 and beyond). In
Section~\ref{sect:redanddead}, we present detailed spectral energy
distributions of massive quiescent galaxies at z$>$1, and demonstrate
the power of our spectro-photometric data to analyze the stellar
populations in this kind of object through a detailed comparison with
stellar population synthesis models.


Throughout this paper we use AB magnitudes. We adopt the cosmology
$H_{0}=70$ km~s$^{-1}$Mpc$^{-1}$, $\Omega_{m}=0.3$, and
$\Omega_{\lambda}=0.7$.


\section{Survey description}
\label{sect:shards_description}

\begin{figure}
  \begin{center}
    \includegraphics[angle=0,width=9.cm]{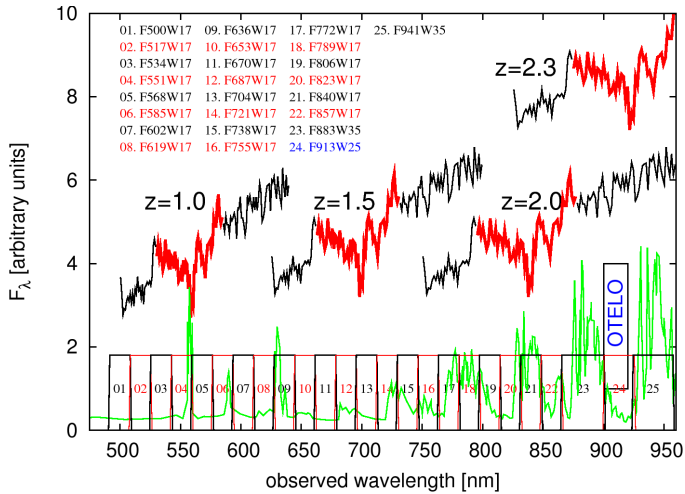}
    \figcaption{\label{fig:allfilters}Observing strategy of SHARDS.
      The figure shows a scheme of the transmission curves for the 24
      SHARDS medium-band filters (black and red lines at the bottom,
      with names giving nominal central wavelengths and widths, as
      written on the legend).  This filter set was designed to probe
      the optical wavelength range contiguously between 500 and 950~nm
      with filters of width FWHM$\sim$17~nm (spectral resolution
      R$\sim$50). The observing strategy of SHARDS was devised to
      identify quiescent galaxies at z$>$1. We show how the stacked
      spectrum of the 13 spectroscopically confirmed quiescent massive
      galaxies at 1.4$<$z$<$2.0 in GOODS-S \citep[][the stack adds up
      480 hours of VLT time]{2008A&A...482...21C} would look like at 4
      different redshifts. The sky spectrum is also depicted in green.
      Using 24 filters, SHARDS probes the prominent absorption feature
      placed at $\lambda$=265-–295~nm, distinctive of passively
      evolving galaxies (marked in red) with a resolution R$\sim$50,
      and it is able to measure its strength with a
      spectro-photometric technique to accurately determine stellar
      age and distances for individual high-z galaxies. The 900~nm
      atmospheric window is not covered by our survey, since it will
      be covered by another approved GTC Guaranteed Time project
      (OTELO, \citealt{2011hsa6.conf..167C,2011PASP..123..252L}).}
  \end{center}
\end{figure}

The Survey for High-z Absorption Red and Dead Sources is a medium-band
optical survey currently being carried out with the Spanish 10.4m
telescope, Gran Telescopio Canarias (GTC), and its OSIRIS instrument.
SHARDS was approved in 2010 as an ESO/GTC Large Program and awarded
with 180 hours of observing time to obtain data through 24 contiguous
medium-band filters covering the wavelength range between 500 and 950
nm. The survey targets the GOODS-N field, covering most of the area
observed with the HST/ACS instrument.  SHARDS was conceived to study
in detail the properties of the stellar populations in 0$<$z$<$4
galaxies (and beyond), focusing in the major goal of analyzing
quiescent massive galaxies at z$=$1.0--2.5. To achieve this goal, the
survey was planned to obtain photometric data in as many filters as
necessary to cover the entire optical window with enough spectral
resolution to be able to measure reliably absorption indices which
could be used to perform a detailed and robust stellar population
synthesis. Indices such as D(4000) or \Mg\, use spectral windows
10-20~nm widths. To obtain this spectral resolution (or better) at
z$>$1, we imposed filter widths of approximately 15--20~nm. A
compromise between spectral resolution, depth, and manufacturing
limitations was adopted and the survey was planned to use 17~nm wide
filters. The bright night sky at wavelengths beyond $\sim$800~nm
imposes that filters as narrow as 17~nm would not reach the needed
magnitude limit to study high-z sources, so our 2 reddest filters are
twice as wide. SHARDS was conceived to reach the typical magnitudes of
sub-L$^*$ galaxies at z$>$1 for every single filter, so the goal was
to obtain a depth of 26.5-27.0~mag at the 3$\sigma$ level with
sub-arcsec seeing. The final manufacturing process resulted in an
average width for our filter set around 15-16~nm, except for the 2
reddest filters, which have a width of 33-35~nm.
Figure~\ref{fig:allfilters} shows the layout of the filter set and the
observational strategy of SHARDS. The characteristics of the filters
are given in Table~\ref{table:filters}, jointly with other details of
the SHARDS data.

\begin{figure}
  \begin{center}
    \includegraphics[angle=0,width=9.cm,bbllx=25,bblly=48,bburx=574,bbury=530]{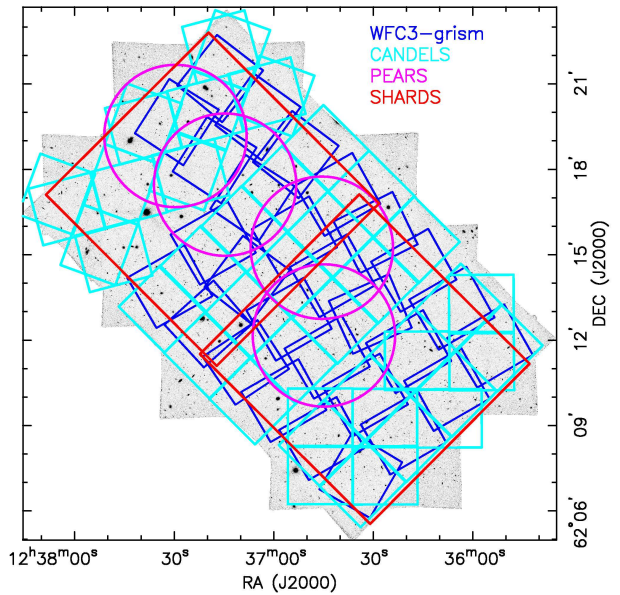}
    \figcaption{\label{fig:footprints}Footprint of the SHARDS data
      overimposed on the ACS images for the GOODS-N field. The
      footprints of the HST grism surveys carried out with ACS (PEARS)
      and WFC3 (PI: Weiner) and the CANDELS coverage of GOODS-N are
      also shown.  SHARDS covers a total surveyed area of
      $\sim$130~arcmin$^2$ divided in two pointings.}
  \end{center}
\end{figure}

SHARDS is being carried out in the GOODS-N field, one of the most
targeted areas of the sky at all wavelengths. Virtually all the deep
region covered by ACS is being surveyed by SHARDS using two OSIRIS
pointings, summing up a total surveyed area of $\sim$130~arcmin$^2$
(see Figure~\ref{fig:footprints}). The central J2000 coordinates of
the two pointings are $\alpha$$=$12:37:18.9 $\delta$$=$$+$62:17:03 and
$\alpha$$=$12:36:33.3 $\delta$$=$$+$62:11:39. A position angle of
45$^\circ$ was used for our imaging data in order to cover the GOODS-N
region more efficiently.

The multi-wavelength dataset available in GOODS-N is extensive,
ranging from an ultra-deep X-ray exposure to the deepest data in the
MIR/FIR with surveys such as FIDEL \citep{2006ApJ...647L...9F}, PEP
\citep{2011A&A...532A..90L}, HerMES \citep{2010A&A...518L..21O}, or
Herschel-GOODS \citep{2011A&A...533A.119E}, as well as multiple
spectroscopic redshifts
\citep{2004AJ....127.3121W,2004AJ....127.3137C, 2006ApJ...653.1004R},
a few of them at z$>$1.5
\citep{2005ApJ...633..748R,2008ApJ...689..687B}.  Very complementary
to the SHARDS data, GOODS-N has been observed by HST with ACS and WFC3
(see Figure~\ref{fig:footprints}) providing slitless, intermediate
resolution spectroscopy in the optical (through the G800L grism;
PEARS, \citealt{2009PASP..121...59K}; see also
\citealt{2004ApJS..154..501P,2009ApJ...695.1591P}) and NIR (G141; PI:
B.  Weiner). GOODS-N is also one of the two fields counting with the
deepest exposures taken by the GOODS \citep{2004ApJ...600L..93G} and
CANDELS \citep{2011ApJS..197...35G,2011ApJS..197...36K} projects. In
addition, the availability of the sky deepest IRAC ([3.6]$<$26.0~mag)
and MIPS [F$_{5\sigma}$(24)$>$30~$\mu$Jy] observations ensures the
detection of the rest-frame NIR/MIR emission of the galaxies and allow
us to robustly estimate stellar masses and SFRs
\citep{2005ApJ...630...82P,2008ApJ...675..234P,2008A&A...482...21C}.
This wealth of data converts GOODS-N in one of the two best fields
(the other being GOODS-S) for the study of the first galaxies and
their evolution, and the best in the northern sky hemisphere.

At the time of publication of this paper, 75\% of the SHARDS data have
already been taken.  Table~\ref{table:filters} presents the data
characteristics for the observed filters in the two SHARDS pointings.

\input{tab1_2c}


\section{Data reduction and calibration}
\label{sect:shards_data}

\subsection{Reduction pipeline}
\label{sect:reduction}

The GTC/OSIRIS instrument \citep{2010hsa5.conf...15C} is an optical
imager equipped with 2 CCDs, which cover a total (usable)
field-of-view of 7.4\arcmin$\times$8.4\arcmin. There is a gap between
the 2 detectors, with size 10--12\arcsec\, as measured directly in our
images in different locations along the gap and after masking
non-useful pixels and correcting for distortions. OSIRIS presents
special characteristics which should be taken into account in the data
reduction procedure. Consequently, apart from the standard reduction
steps (bias subtraction and flat-fielding), our custom OSIRIS pipeline
includes the following additional steps: illumination correction,
background gradient subtraction, fringing removal, WCS alignment
taking into account field distortions, and 2-dimensional calibration
of the photometric pass-band and zeropoint. In what follows, we
elaborate on these steps and describe in detail the non-standard parts
of the reduction.

Bias frames were taken each night. These calibration data were found
to be very stable and uniform, with typical variations of less than
3\% from pixel to pixel and from night to night. Dark current was
found to be negligible in all our data.

Sky flats for each filter in the SHARDS dataset were taken at sunset
and dawn every night. These flats presented strong gradients across
the field (typically, 10-50\% differences in brightness from one edge
of the FOV to another), with significant spatial and temporal
brightness variations. These gradients were also seen in science data
during both dark and (especially) gray night time.

After analyzing night and day-time data, we concluded that these
spatial variations were mainly due to the special characteristics of
the OSIRIS instrument, which operates off-axis for our observational
setup (medium- and broad-band imaging). Indeed, light rays reach the
OSIRIS detector (when not using the tunable filters) in a wide range
of angles of incidence (AOI):
--2$^\circ$$\lesssim$AOI$\lesssim$22$^\circ$.  Given the typical
dependence with the light ray's AOI of the central wavelength (CWL) of
the passband for interference filters, and the medium-width of our
filter set, our flat-field frames were subject to the spectral
features of the sky spectrum, and their temporal variations.

The optical axis is outside the OSIRIS FOV when using a single broad-
or medium-band filter. For this reason, a gradient is observed in our
data (flat-field and science images) with a symmetry around an
horizontal line approximately dissecting in equal parts the FOV in the
vertical direction. The gradient follows a radial profile centered in
the optical axis. It presents varying structures of different
brightnesses whose position and strength depend on time and on the
filter.  The brightest structures are located where strong sky
emission lines (or bands) go through the pass-band for each physical
filter as its CWL is changing along the FOV.

This effect directly related to the special characteristics of OSIRIS
and our instrumental setup means that it is very difficult to find a
spatially constant light source to take flat-fields, i.e., a
significant part of the structure of the flat-field frames is linked
to the sky spectrum.  Thus, our pipeline included an illumination
correction to get rid of this effect as much as possible. This
illumination correction was carried out by comparing the flat-field
images taken for our medium-band filters with those acquired through
broad-band pass-bands (typically, $r$- and $i$-band filters). For
these broad-band filters, the sky spectrum is averaged in a spectral
range which is wide enough to prevent strong spatial variations, and
are also very constant in time. Indeed, the super-skyflat provided by
the observatory and built with thousands of frames shows variations of
less than 5\% along the FOV and from night to night.

We used the ratio of our medium-band flat-fields to the broad-band
flat-fields to correct the SHARDS data for the illumination effect.
These ratios were smoothed using a 3$^\mathrm{th}$-order spline.

After applying the flat-field and illumination corrections to the
science data, these images presented a highly symmetric sky gradient,
which was subtracted with a median filtering and spline interpolation,
after masking objects. In order to avoid the effect of the wings of
the objects in this sky determination, we increased the extension of
the sources (typically by a factor of 2 in Kron radius,
\citealt{1980ApJS...43..305K}) taking into account their brightnesses,
and we also masked out faint objects detected in a preliminary mosaic
constructed for each filter adding all the available observations.

The data for the filters whose CWLs are redder than $\sim$700~nm
presented some fringing (typically with an intensity below 1\% of the
background). We removed this additive effect using the {\sc rmfringe}
task in IRAF\footnote{IRAF is distributed by the National Optical
  Astronomy Observatory, which is operated by the Association of
  Universities for Research in Astronomy (AURA) under cooperative
  agreement with the National Science Foundation.}.

Before stacking all the data together, we calibrated the World
Coordinate System (WCS) for each image using the positions of several
hundred objects for each CCD, which were cross-correlated with a
sample of galaxies detected in the Subaru $R$-band image of GOODS-N.
For this task, we used the SCamp program \citep{2006ASPC..351..112B},
and the WCS utilities in IRAF, obtaining for each frame an undistorted
image remapped to a TAN-SIP coordinate system. The distortion in
individual frames reaches several arcseconds in the edges, and the
final WCS calibrated images present a typical position uncertainty
around 0.1\arcsec\, throughout the FOV.

Finally, the SHARDS pipeline stacks together all the data for a given
filter with a sigma-clipping algorithm to get rid of cosmic rays and
artifacts, producing final mosaics and exposure maps with the same WCS
for all SHARDS filters and for each of our 2 pointings covering the
GOODS-N field.

\subsection{Calibration procedures}

Given the special characteristics of the OSIRIS instrument at GTC,
within each single frame taken with a given physical filter, each
pixel sees a different pass-band. This has a strong effect on the
images taken with medium-band filters, where the shift of the central
wavelength of the actual pass-band seen by different parts of the
detector produces sky gradients and significant difficulties in
obtaining flat-field calibration images, as explained in the previous
section. To overcome this issue, we performed a detailed calibration
of the effective pass-band and photometric zero-point as a function of
the position in the FOV. Although part of the spatial variation of the
photometric zeropoint can be removed by applying an illumination
correction built with broad-band flat-field images, a small effect can
remain in the images. To account for this, during our calibration
procedure we considered that the zeropoint could vary along the image.
The calibration method was devised, consequently, to determine the
spatial variation of the zeropoint.  Note that after the calibration
routine, we can construct a photometric catalog for each physical
filter, but within this catalog, the flux measurements for each
detected source refer to a different pass-band.  This is quite
different from a standard photometric catalog for a given physical
filter, where all sources share the same pass-band.

We describe the procedures to carry out the calibration of the SHARDS
data in the following subsections, starting from the CWL calibration
and following with the absolute flux calibration.

\subsection{Pass-band central wavelength (CWL) calibration}
\label{sect:cwl}

\begin{figure}
  \begin{center}
    \includegraphics[angle=-90,width=8.cm]{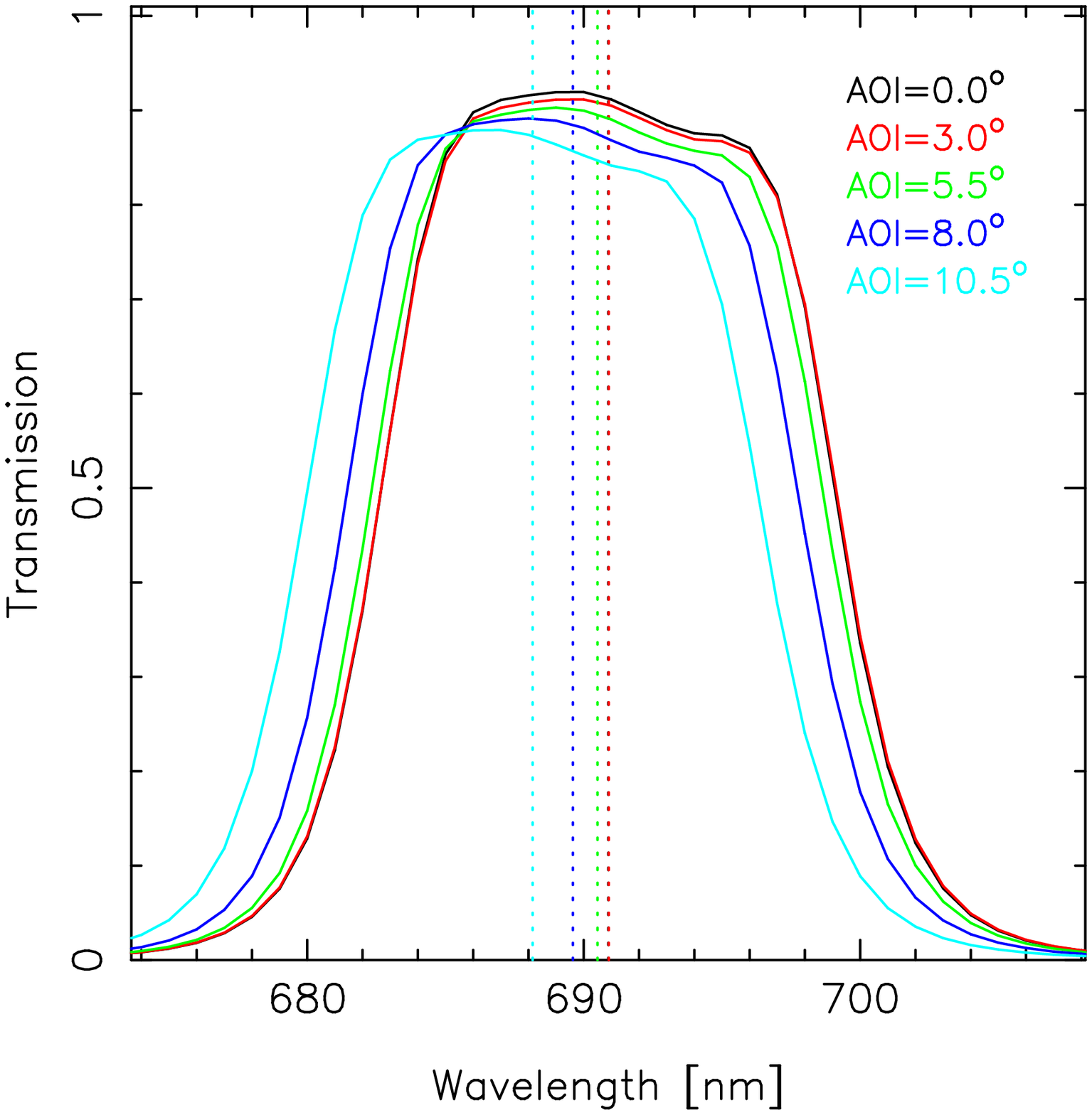}
    \includegraphics[angle=-90,width=8.cm]{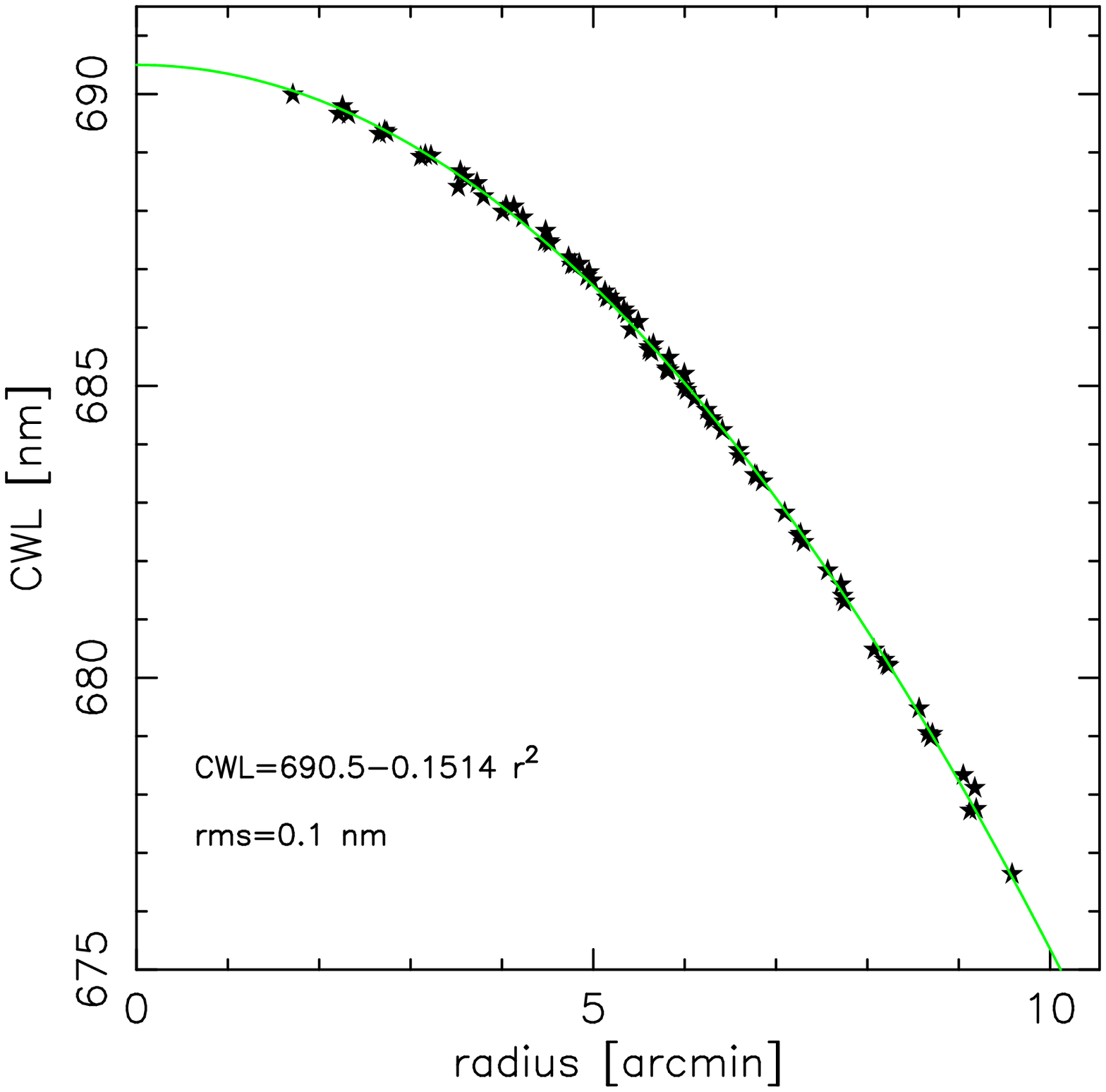}
    \figcaption{\label{fig:cwl_calib}{\it Top:} Calibration of the
      transmission curve of the F687W17 filter as a function of
      angle-of-incidence, obtained with laboratory data. The central
      wavelengths of each pass-band are marked with vertical lines.
      {\it Bottom:} Calibration of the spatial variation of the CWL of
      the F687W17 filter along the OSIRIS FOV.  This variation is
      symmetric and only depends on the distance to the optical axis.
      The data are fitted with the function in Equation~\ref{equ:cwl}
      (green line), leaving the position of the optical axis as a
      fitting parameter. The best fit between CWL and distance to the
      optical axis in arcmin is given in the plot, jointly with its
      rms. The width of the pass-bands is relatively independent
      (differences within 0.2~nm) of the position in the FOV. }
  \end{center}
\end{figure}

The central wavelength of the pass-band seen by different parts of the
OSIRIS detector for a given physical filter varies along the FOV. This
dependence was calibrated with day-time imaging and spectroscopic data
taken in laboratory and at GTC. In the lab, we calibrated the
transmission curve for each filter as a function of the
angle-of-incidence of the incoming beam. At the telescope, once the
given filter was mounted in OSIRIS, we took spectroscopic data through
a special mask with 105 pinholes homogeneously covering the entire
FOV.  The positions of those pinholes were determined with images
taken through the mask and using one of our filters. Then, we also
took spectra of these pinholes with the R1000B and R1000R grisms
(which cover the entire wavelength range probed by SHARDS) through our
filters, calibrating them with Ne, HgAr, and Xe arcs. Typical
uncertainties in the wavelength solution were smaller than 0.01~nm.
These data were used to measure the transmission curve at each pinhole
position. We characterized each curve using two parameters: the CWL
and the width of the pass-band. The shape and width of the pass-bands
are relatively independent of the position in the FOV (i.e., all
pinholes show similar values within 0.2~nm). However, the CWL presents
a significant symmetric variation around the optical axis, located to
the left of the FOV.  The CWLs for all the pinholes were fitted with a
function depending on the square of the distance to the optical axis
(r$^2$), leaving also the position of the optical axis as a parameter
to fit (see \citealt{2011PASP..123.1107M} and
\citealt{2012arXiv1203.1842M} for similar calibration procedure, but
for the OSIRIS red tunable filter).  An example of the calibration for
one of the SHARDS filters is shown in Figure~\ref{fig:cwl_calib}. The
data were fitted with the following function:

\begin{equation}
\label{equ:cwl}
CWL(X,Y)=A+B\times[{(X-X_0)^2+(Y-Y_0)^2}]
\end{equation}

\noindent where X and Y are the positions (in pixels) in the OSIRIS
FOV, and $X_0$ and $Y_0$ are the position of the optical axis,
combined all of them to give a distance to the optical axis in pixels.
$X_0$, $Y_0$, $A$, and $B$ were measured for each one of the SHARDS
filters by fitting the data described above and shown in
Figure~\ref{fig:cwl_calib} for filter F687W17. When applied to the
actual science data, the optical axis position was converted to RA and
DEC, and the distances were measured in arcmin using the nominal pixel
size of our data, 0.251 arcsec/pixel.

The fitting coefficients in Equation~\ref{equ:cwl} for each one of the
SHARDS filters (with available data so far) are given in
Table~\ref{table:filters}. We want to remark that the four parameters
in Equation~\ref{equ:cwl} are highly correlated. Consequently, the
differences seen from filter to filter may not be closely related to
real differences in the position of the optical axis or CWL variation
coefficients. In any case, we are only interested in recovering CWL
values for any position within the FOV, and even assuming large
(correlated) uncertainties for the four parameters, the results given
in Table~\ref{table:filters} do provide very accurate CWLs. In
addition, we tested the repeatability of these measurements and no
significant changes in this calibration were detected for observations
taken in different nights.


\subsection{Photometric calibration}

\begin{figure*}
  \begin{center}
    \includegraphics[angle=-90,width=10.cm]{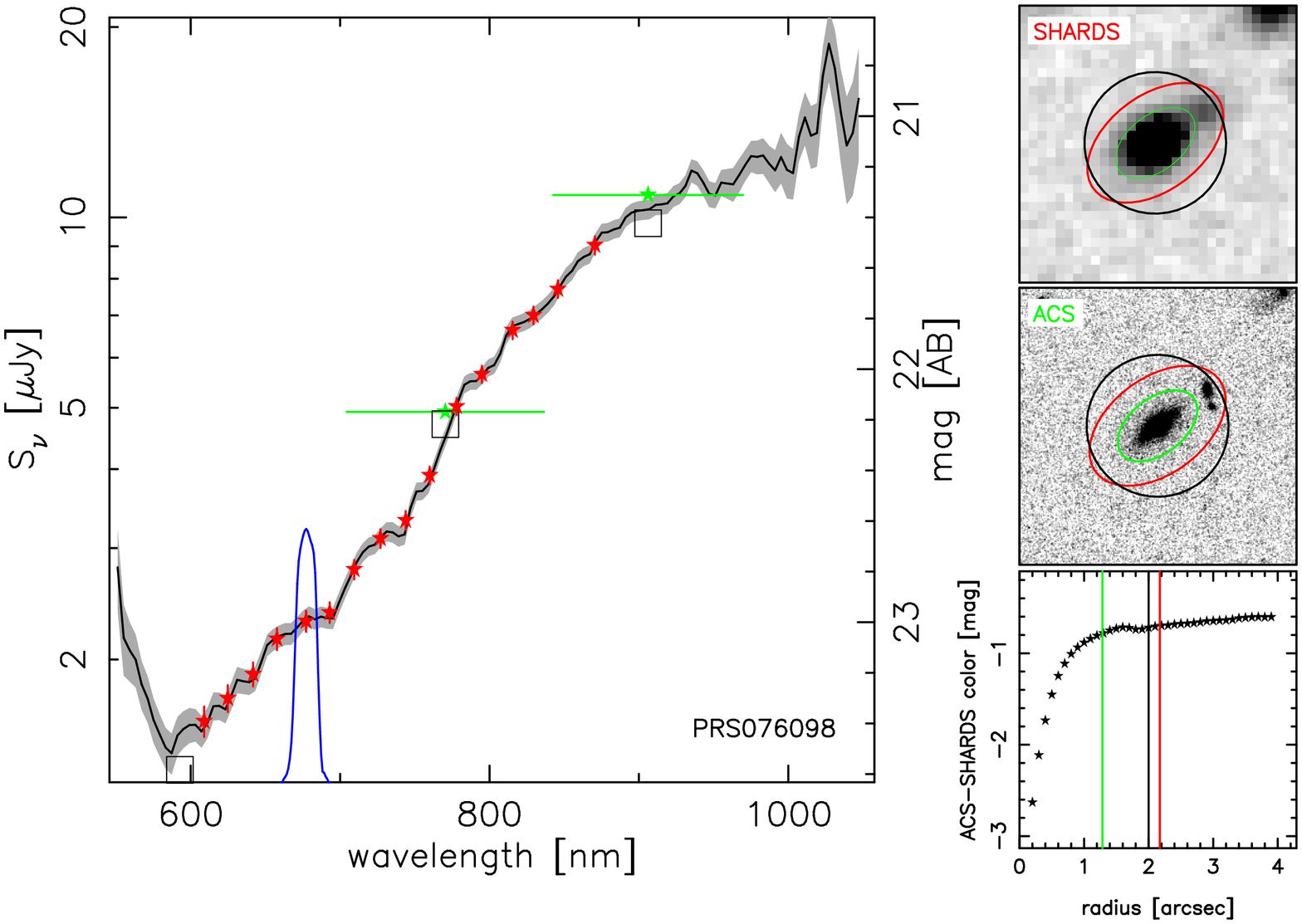}
    \includegraphics[angle=-90,width=10.cm]{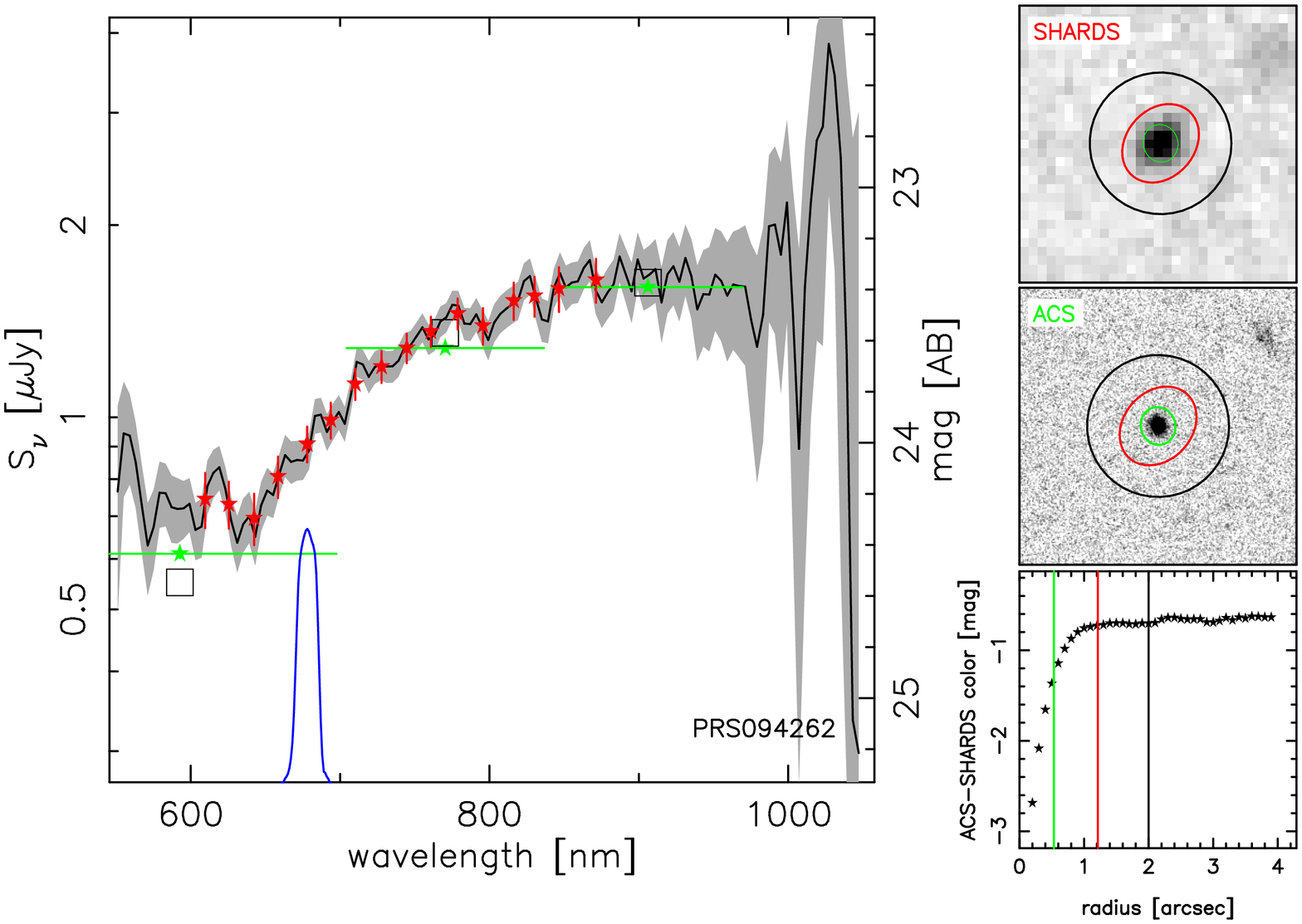}
    \includegraphics[angle=-90,width=10.cm]{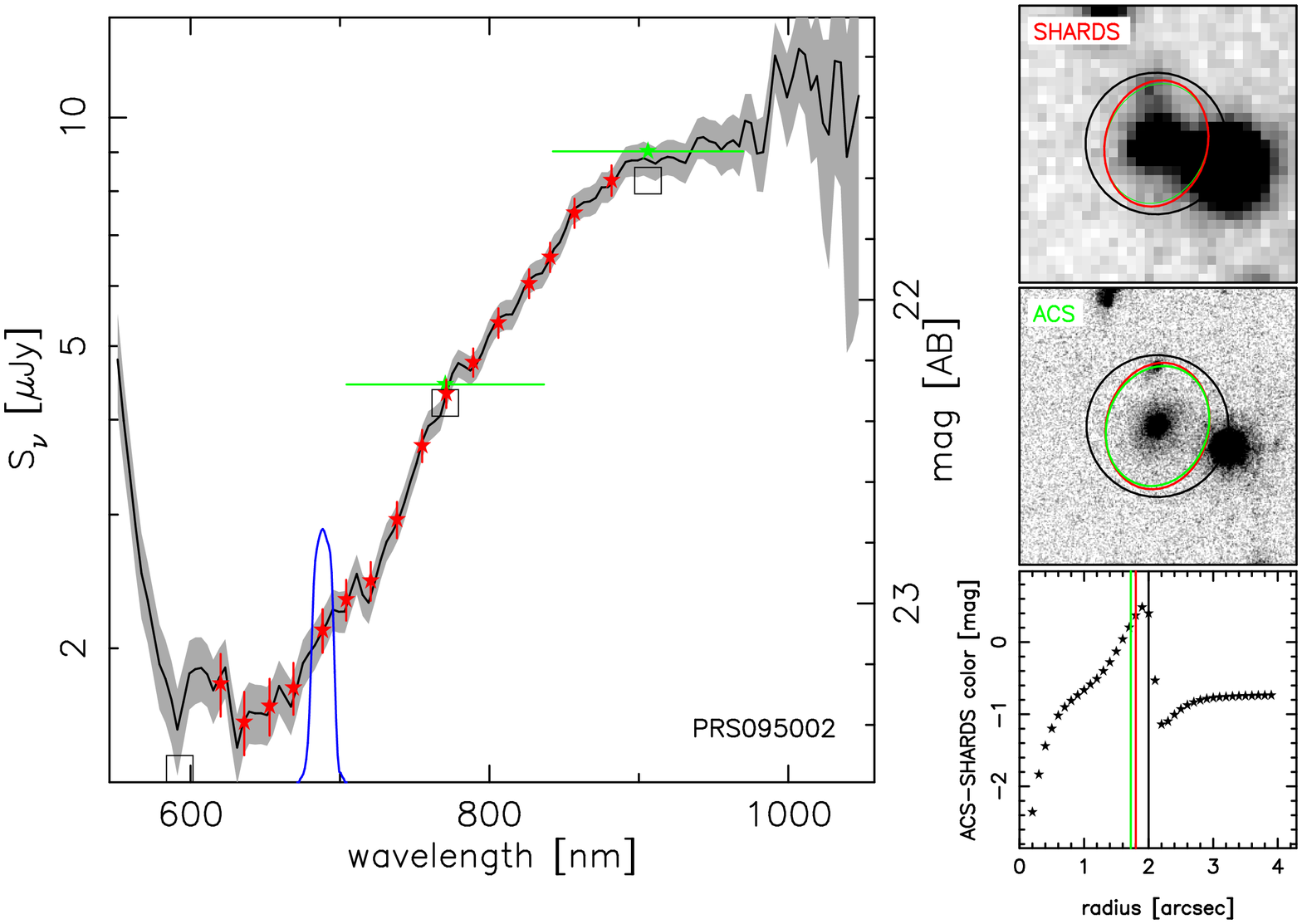}
    \figcaption{\label{fig:acscalib} Examples of the HST/ACS spectra
      (from the PEARS data described in \citealt{2009ApJ...706..158F}
      and the PEARS database) used in the calibration of the SHARDS
      data. The ACS spectra are shown in black with the gray shaded
      area depicting the uncertainties. The spectra are scaled to the
      ACS $viz$ photometry, shown in green, including also
      uncertainties and filter widths. The convolutions of the spectra
      with the $viz$ filter transmission curves are shown with
      squares.  The pass-band for the F687W17 filter seen by each
      galaxy is shown in blue, and the convolution with the HST
      spectra is marked with a red star, jointly with the convolutions
      for the other SHARDS filters observed so far (no transmission
      curves are depicted for the sake of clarity). We show the SHARDS
      and $i$-band ACS images of the galaxy
      (5$\arcsec$$\times$5$\arcsec$) and the color between them as a
      function of aperture radius. In each plot, the best elliptical
      aperture determined by sextractor \citep{1996A&AS..117..393B}
      for the ACS data is shown in green, the best aperture for the
      SHARDS image is plotted in red, and the circular aperture used
      to calibrate the SHARDS data is shown in black. The panel in the
      last row shows a galaxy whose photometry (and probably the
      spectra) are affected by contamination from nearby sources.
      This kind of source was excluded from our calibration procedure.
      They are marked with gray symbols in
      Figure~\ref{fig:acsoffset}.}
  \end{center}
\end{figure*}

The significant variation of the passband seen by each point of the
detector as a function of the position in the FOV implies a complex
behavior of the absolute photometric calibration of the SHARDS images.
Moreover, the effects of this CWL variation on the construction of the
flat-field (explained in Section~\ref{sect:reduction}) may also affect
the flux calibration of the final mosaics (see
\citealt{2012arXiv1203.1842M} for a description of the same problem,
but for the OSIRIS tunable filters). To cope with these issues, we
developed a special flux calibration procedure, aimed at determining
the zeropoint of the SHARDS mosaics in each filter as a function of
position in the image. Note that the behavior of the passband is
symmetric around the optical axis and each position in the FOV is
characterized by a central wavelength (as explained in
Section~\ref{sect:cwl} and defined by Equation~\ref{equ:cwl}). Keeping
this in mind, we work with zeropoint variations as a function of CWL,
instead of directly relating the calibration to a position in the
images.

The flux calibration of the SHARDS mosaics was performed by comparing
the measured photometry in our images with spectroscopic data for
several sources in the field (most of them being galaxies). Given the
need for spectroscopic calibration data for sources covering the whole
FOV, direct observations taken with GTC/OSIRIS were unaffordably
expensive in terms of observing time, and even impossible since OSIRIS
currently has limited spectroscopic capabilities (only long-slit). We
therefore used spectroscopic data taken with other telescopes. In any
case, a spectro-photometric standard star was observed with OSIRIS
with a grism and each one of our filters, and we checked that our main
calibration procedure (described below) is consistent with these (very
limited) observations.

For our main calibration procedure, synthetic fluxes were obtained by
convolving the spectra for sources distributed all along the FOV with
the appropriate transmission curve seen by that source, according to
our calibration of the passband (in terms of the position in the FOV,
see Section~\ref{sect:cwl}).  The spectra used in this calibration
method were taken from 2 different sources: (1) the HST ACS grism data
from the PEARS project \footnote{Downloaded from the database at
  http://archive.stsci.edu/prepds/pears/. See also
  \citet{2009ApJ...706..158F}.}; and (2) the spectroscopic data
released by the Team Keck Treasury Redshift Survey
\citep[TKRS][]{2004AJ....127.3121W} and DEEP3
\citep{2011ApJS..193...14C} taken with the Keck/DEIMOS instrument. We
note that the TKRS and DEEP3 spectra are not flux calibrated, they are
not completely flat-fielded in the spectral direction (see below), and
are subject to the effect of strong sky emission lines and telluric
absorption bands in certain spectral regions. For these reasons, our
primary calibrator was the PEARS dataset of HST/ACS spectra.  The Keck
spectra were used as a consistency check for the primary calibration.
In addition, we also performed another test of the calibration based
on synthetic magnitudes obtained from stellar population models
fitting the broad-band photometry for the galaxies with a published
spectroscopic redshift and compiled in the Rainbow Cosmological
Surveys database
\citep[see][]{2008ApJ...675..234P,2011ApJS..193...13B,2011ApJS..193...30B}.
A finer calibration will be carried out in the future based on high
SNR (SNR$>$10) spectra for bright objects which are currently being
obtained at TNG and the CAHA 3.5m telescope. The several methods to
calibrate the data are explained in the following sub-sections.

\subsubsection{Primary calibrator: HST/ACS grism spectroscopic data}

The photometric calibration procedure starts with a crude
determination of the zeropoint based on the comparison of photometry
in the SHARDS images with $R$- or $I$-band fluxes. This comparison
provides a first estimation of the conversion from counts to AB
magnitudes (one value for a given image). Then, we use the HST/ACS
spectra to improve this calibration and check the dependence on the
position along the FOV. We start by scaling the ACS spectra to the
broad-band photometry in the $viz$ bands from ACS. The rms of the
comparison of broad-band photometry and the synthetic fluxes obtained
with the grism spectra is 0.38~mag, i.e., this is the typical accuracy
of the spectroscopic data for each galaxy. The flux measurement in the
ACS bands is carried out within a large enough aperture so it can be
compared with photometry in the SHARDS images with negligible seeing
effects and also avoiding (as much as possible) the contamination from
nearby sources.  Typically, apertures with radii larger than
1$\arcsec$ were used. The procedure basics are outlined in
Figure~\ref{fig:acscalib}: the spectra are scaled to the broad-band
photometry, we measure the flux in the SHARDS pass-band corresponding
to the position of the galaxy in the FOV by convolving the spectra
with the appropriate transmission curve, and we compare with the flux
in counts measured in the SHARDS image, obtaining a conversion from
counts to flux densities or AB magnitudes as a function of position in
the FOV.

\begin{figure}
  \begin{center}
    \includegraphics[angle=-90,width=8cm]{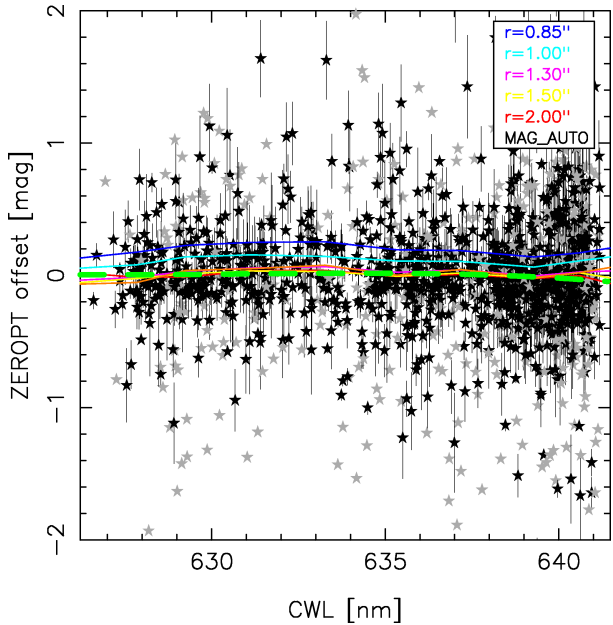}
    \figcaption{\label{fig:acsoffset} Calibration of the SHARDS data
      for filter F636W17 based on the HST/ACS spectra. The difference
      between the AB magnitude obtained from the spectra and the
      magnitude measured in the SHARDS images is plotted as a function
      of CWL of the pass-band seen by the galaxy (i.e., vs.  position
      in the field-of-view, as described in Section~\ref{sect:cwl} and
      defined by Equation~\ref{equ:cwl}).  The points show the
      calibration obtained from individual galaxies for the best
      photometric aperture, i.e., that enclosing the entire object and
      minimizing the uncertainties, the effects of seeing, and the
      contamination from close neighbors. Error bars are estimated
      from the grism flux uncertainties and the rms of the comparison
      of ACS spectra and broad-band photometry in the $bviz$ filters
      (0.4~mag).  Objects whose photometry is affected by nearby
      sources or have low-SNR spectra are plotted with gray symbols.
      Color solid lines show the median behavior for the offset
      calculated with different photometric aperture radii. The final
      calibration is shown with a green dashed line, resulting from a
      polynomial fit to the best aperture data.}
  \end{center}
\end{figure}

\begin{figure*}
  \begin{center}
    \includegraphics[angle=-90,width=16cm]{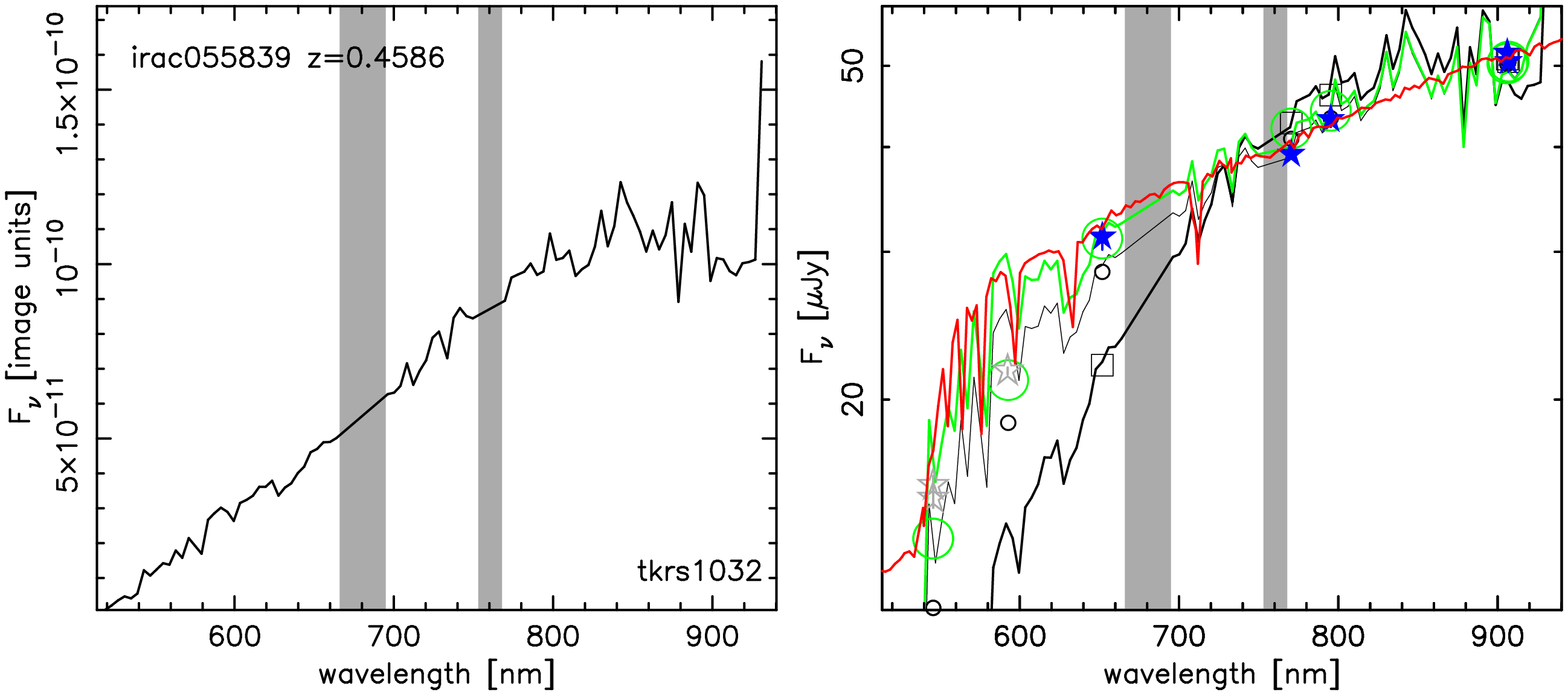}
    \includegraphics[angle=-90,width=16cm]{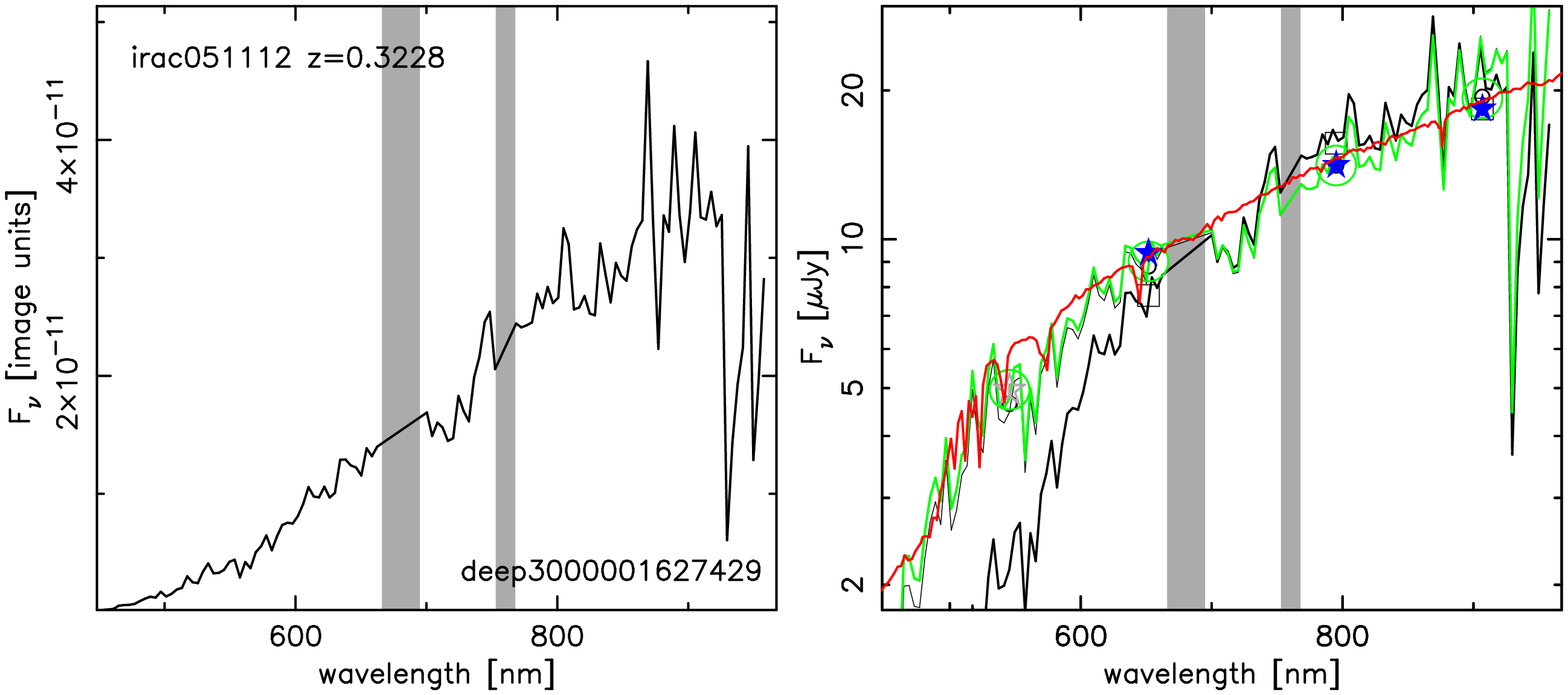}
    \figcaption{\label{fig:speckeck} Two examples of the ground-based
      spectra used to test the absolute flux calibration of the SHARDS
      data. For each galaxy, the original spectrum (in arbitrary
      F$_\nu$ units and binned to achieve a resolution of 2~nm) is
      plotted on the left, indicating the name of the source in the
      Rainbow Cosmological Surveys database
      \citep{2008ApJ...675..234P} the TKRS and DEEP3 catalogs
      \citep{2004AJ....127.3121W,2011ApJS..193...14C} and the
      spectroscopic redshift. Both on the left and right panels, we
      mark with a shaded area the regions where strong telluric
      absorption bands are located. On the right, we show with a black
      thick line the original spectrum after flux scaling it to match
      the broad-band photometry, depicted with blue stars (only the
      reddest points were used in this first scaling, given that they
      provided a better absolute calibration) and gray symbols (for
      the bluest bands which are more affected by the spectral flat
      and atmospheric extinction corrections).  The black thin line
      shows the same spectrum after applying an average spectral flat
      obtained as explained in the text. The green line shows the
      final ground-based spectrum used for the calibration of SHARDS
      data, obtained after applying an atmospheric extinction
      correction calculated through a comparison with a stellar
      population synthesis model fitting the broad-band photometric
      data, shown in red in the plots. Black squares and circles, and
      green circles show the convolution of the previously described
      spectra (respectively: the original after flux calibration, the
      one after further calibration with a spectral flat-fielding, and
      the final one after carrying out also an extinction correction)
      with the transmission curves for the broad-band filters.}
  \end{center}
\end{figure*}

Figure~\ref{fig:acsoffset} shows the results of the calibration based
on HST/ACS spectra. We plot the offset between the calibration
obtained from the analysis of the HST spectra and the preliminary
calibration obtained with the comparison with $R$-band fluxes as a
function of the CWL of the passband seen by each galaxy. In this
example, we show the results for the observations of one of the SHARDS
pointings through the F636W17 filter, where we have more than 1,500
ACS spectra to compare with. For the second pointing, PEARS only
covered a part of it, and the number of available ACS spectra is
$\sim$900. The derived differential calibration as a function of
wavelength (i.e., position) was applied to the images, so the final
mosaics have a constant zero-point throughout the whole image. The
typical scatter around the average calibration was below 0.1~mag.
Note that many spectra provided very large zeropoint offsets. We
visually inspected all spectra with offsets above 0.5~mag and
virtually all of them presented nearby bright objects which were most
probably contaminating the spectra (e.g., the last source in
Figure~\ref{fig:acscalib}).  This effect was confirmed by comparing
the scale factor of the spectra to the broad-band photometry. In
spectra affected by contamination by nearby objects, the three bands
used in the scaling provided very different factors, with the scatter
being considerably larger than the typical value ($\sim$0.4~mag). Note
also that we did not establish any magnitude cut in the HST/ACS data,
and a significant fraction of the spectra correspond to very faint
objects and count with large uncertainties.

\subsubsection{Calibration test: ground-based spectroscopic data}

The calibration obtained with the HST data was checked with other
spectra taken from the literature. We found nearly 1,000 spectra for
sources detected by SHARDS in the TKRS database and the DEEP3 release
in GOODS-N \citep{2004AJ....127.3121W,2011ApJS..193...14C}.  The
spectra were all taken with the DEIMOS instrument on Keck with typical
exposure times of 1-2 hours through R$\sim$300--600 grisms.  Given
that these surveys were mainly interested in spectroscopic redshifts
based on emission-lines, the typical SNR in the continuum is low
(SNR$\sim$1 at the original spectral resolution). In addition, the
spectra are not flux calibrated and they typically show a decrease of
flux to the blue that seems to arise from the lack of an accurate
spectral flat correction (see Figure~\ref{fig:speckeck}). Moreover,
these spectra were taken through a mask with slit widths of 1\arcsec,
so aperture effects may also be significant, i.e., the spectroscopic
data give information about the central part of the galaxies and a
comparison with seeing limited photometry can be biased. For these
reasons, we just used these spectra as a test for our basic absolute
flux calibration based on HST data. Prior to this test, we had to
correct the individual ground-based spectra with an average spectral
flat and flux calibrate them. Both procedures were carried out using
broad-band photometry and fits to this photometry with stellar
population synthesis models from the Rainbow Cosmological Surveys
database \citep{2008ApJ...675..234P}.

Figure~\ref{fig:speckeck} shows two typical spectra in GOODS-N
extracted from the ground-based spectroscopic sample. After binning
the spectrum to a resolution of 2~nm, we carried out a preliminary
flux calibration of the spectrum based on the photometric broad-band
data around 800~nm (blue stars in the right panels of the figure). An
inspection of the resulting calibrated spectra (thick black lines in
Figure~\ref{fig:speckeck}) revealed that the spectroscopic data were
not corrected with an adequate spectral flat, resulting in a large and
increasing flux difference between the spectra and the broad-band
photometric data at shorter wavelengths (see comparison between black
thick lines and the photometry in Figure~\ref{fig:speckeck}). To
correct for this effect, we built a master spectral flat by averaging
the ratio between the spectra and stellar population synthesis models
fitting the broad-band photometry (red lines in
Figure~\ref{fig:speckeck}) for all sources in the spectroscopic
sample. After applying this correction, the comparison between the
spectra (thin black lines in Figure~\ref{fig:speckeck}) and the
photometry was better, but still most sources presented a dimmer flux
level in the spectra when compared with the photometry for the bluest
wavelengths.  We identified this flux difference as an effect of
differential extinction, which was not fully taken into account in the
average spectral flat. To account for this, we applied a final
correction based on a typical extinction curve at Mauna Kea scaled to
explain the difference between the spectra and the broad-band
photometry (especially in the bluest bands). This final spectrum
(green lines in Figure~\ref{fig:speckeck}) was then convolved with the
SHARDS filter pass-band (the appropriate one according to the position
of the galaxy in the FOV), providing another independent calibration
for our data.

\subsubsection{Calibration test: synthetic photometry}

Finally, the fluxes obtained in the SHARDS bands were compared
directly with synthetic magnitudes obtained by convolving the SHARDS
filter pass-bands with stellar population synthesis models fitting the
broad-band photometry for each source. Note that this calibration test
is not completely independent of the one carried out with the
ground-based spectra since those spectra were partially calibrated
with the stellar population fits. Moreover, the comparison with models
for individual galaxies may be biased due to the presence of
emission-lines, which are not constrained by the broad-band
photometry. However, because of the lower noise of the templates, we
can compare them with a large number of galaxies. Moreover, even the
fits for galaxies with no spectroscopic redshift could be reliable in
a statistical way because our convolutions with the SHARDS filters are
interpolations between broad-band fluxes. Therefore, the comparison
with the templates is another good test of the absolute flux
calibration, and it indeed provided reassuring results for all
filters.

In summary, the flux calibration of the SHARDS data was primarily
based on the comparison with HST/ACS grism spectra. This calibration
was tested through a comparison with ground-based spectra and stellar
population synthesis models fitting the broad-band photometry. Based
on these three different calibrators, we also estimated the typical
uncertainty in the zeropoints of the SHARDS images, typically
0.05--0.08~mag (given in Table~\ref{table:filters}).

\subsection{Data quality}

The SHARDS data presented in this paper were obtained in queue mode,
and a maximum seeing threshold was set to 0.9\arcsec. This image
quality was imposed in order to reach the necessary flux limits to
detect and reliably measure absorption indices for massive quiescent
galaxies up to z$\sim$2.5 (26.5-27.0~mag at the 3$\sigma$ level). In
Table~\ref{table:filters}, we show the final seeing of the SHARDS
images, measured after mosaicking all the data for a given filter and
carrying out all reduction and calibration steps. Virtually all SHARDS
data were taken under sub-arcsec conditions.

Concerning the final depth of our images, we give in
Table~\ref{table:filters} the typical depths reached at the 3$\sigma$
level (measured in sextractor-based mag\_auto apertures), and the
magnitude level corresponding to the third quartile of the brightness
distribution.  The desired depths were reached in most of the filters,
with some of them presenting exceptional seeing and depth figures.


\section{SHARDS Science verification: emission-line galaxies}

The analysis of emission lines in the spectra of star-forming galaxies
(SFGs) and objects hosting AGN is one of the main tools to understand
galaxy evolution. Emission lines can be used to select both SFGs and
AGN, and then to obtain estimations of relevant physical parameters,
being redshift, SFR, metallicity, or black hole mass some of the most
interesting \citep[see,
e.g.,][]{2001MNRAS.323..887C,2002ApJS..142...35K,2010MNRAS.405.2594G,
  2012MNRAS.422..215Y}.

Although the most straight-forward way of getting emission-line
identifications and fluxes is through spectroscopic observations,
these data are hard to obtain for large number of objects and very
time consuming. Moreover, current spectrographs on the largest
telescopes are only able to reach continuum magnitudes around
$RI$$\sim$24-25~mag. This results in a scarcity of spectroscopic data
for faint objects, especially at z$\gtrsim$1.5, and a bias of
spectroscopic surveys towards bright emission-line galaxies at
z$\lesssim$1.

A powerful alternative to spectroscopic surveys can be found in
(ultra-)deep narrow-band imaging observations, which have been
demonstrated to be useful to select emission-line galaxies (ELGs) with
the faintest magnitudes, and measure important quantities such as
equivalent widths (EWs) and line fluxes \citep[see, among
many,][]{1998ApJ...506..519T,2003PASJ...55L..17K,2001A&A...379..798P,
  2004ApJ...611..660O,2005MNRAS.357.1348W,2007ApJ...657..738L,
  2007ApJS..172..456T,2008ApJ...677..169V,2011ApJ...740...47V,
  2008ApJS..175..128S, 2009A&A...498...13N,2009MNRAS.398...75S,
  2010ApJ...714..255G,2010Natur.464..562H,2011ApJ...726..109L}.  In
addition, although imaging data cannot directly provide robust
identifications and precise observed wavelengths for emission-lines,
they certainly help determine accurate photometric redshifts, even for
very high redshift sources, based on the detection of both emission
and absorption features or breaks.

\subsection{ELGs at intermediate redshift}
\label{sect:elgs}

In this section, we analyze the sensitivity and spectral resolution of
our ultra-deep medium-band survey to select ELGs and measure their
relevant parameters (EW, flux).  Imaging surveys in the optical aimed
at selecting ELGs usually employ narrow-band filters, with narrow
meaning widths around 10~nm. The SHARDS filters are wider, the typical
FWHM is 15~nm, but the depth, photometric accuracy, and imaging
quality of the GTC data (see Table~\ref{table:filters}) can compensate
the lower spectral resolution (R$\sim$50), compared to more classical
narrow-band surveys.

\begin{figure}
  \begin{center}
    \includegraphics[angle=-90,width=8.5cm]{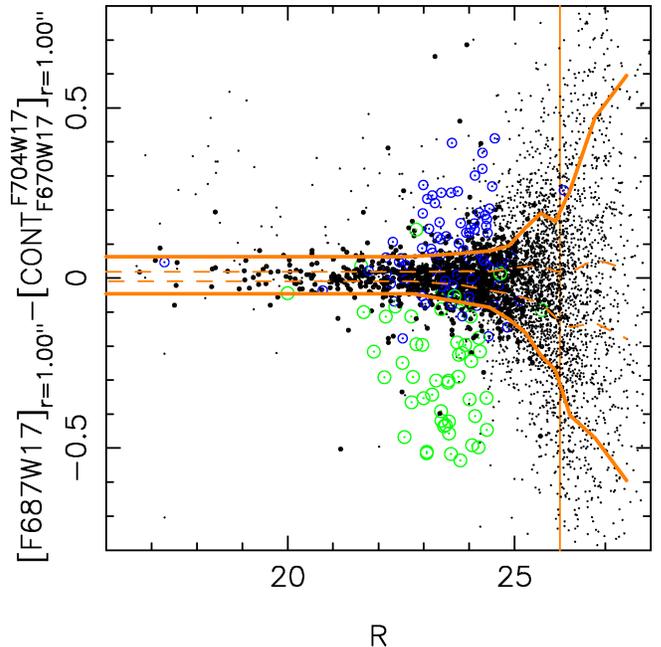}
    \figcaption{\label{fig:trumpet} Color-magnitude diagram showing
      emission-line galaxy candidates at $\sim$687~nm in SHARDS
      pointing \#1. The vertical axis shows the color between the
      F687W17 band and the average magnitude in the adjacent SHARDS
      bands (F670W17 and F704W17) measured within a circular aperture
      of radius r$=$1\arcsec. Sources with measured spectroscopic
      redshifts implying that the [OII] line lies within the F687W17
      filter are marked in green, while sources whose [OII] emission
      is located in the adjacent filters are marked in blue. Solid
      points show galaxies with a photo-z between 0.7$<$z$<$0.9. The
      dashed orange lines depict the typical photometric uncertainty
      as a function of magnitude. The locus for ELGs with an emission
      line within the F687W17 filter detected with more than
      3-$\sigma$ confidence is the region below the bottom orange
      continuous line.  ELGs with emission-lines within the F670W17 or
      F704W17 pass-bands should be located above the top orange
      continuous line. The vertical orange line shows the 5-$\sigma$
      detection threshold of the SHARDS survey in the F687W17 band.
      Only the sources detected by IRAC for which robust photometric
      redshifts were estimated \citep{2008ApJ...675..234P} are
      depicted.}
  \end{center}
\end{figure}

Figure~\ref{fig:trumpet} shows an example of the selection of ELGs
with SHARDS data for one of our 24 filters, the one centered at 687~nm
(filter F687W17).  The method is similar to that used by narrow-band
surveys such as the ones referenced above: the flux in a given
photometric band at wavelength $\lambda_\mathrm{line}$ is compared
with the average flux around that wavelength. Sources with
emission-lines lying inside the central filter would present an excess
of flux compared to the average around it, with the latter
corresponding to the spectrum continuum next to the line.  In this
particular example, we would expect to identify a population of
galaxies at z$\sim$0.80-0.84 featuring an excess in the flux seen by
the F687W17 filter due to [OII] emission.  Note that the redshift
interval is governed by the width of the filter as well as by the CWL
variation of the pass-band along the FOV.  Other lines could provide
samples at different redshifts (e.g., Ly-$\alpha$ at z$\sim$4.6).

Typical narrow-band surveys use a broad-band filter to determine the
continuum
\citep[e.g.,][]{2008ApJ...677..169V,2007ApJS..172..456T,2009MNRAS.398...75S,
  2010A&A...509L...5H,2012MNRAS.420.1926S}, or one or several narrow-
or medium-band filters around a given one \citep{2012arXiv1205.0017L}.
In our case, the contiguous spectral coverage of the optical window
allows us to obtain a continuum determination by using the adjacent
filters to a given one, or even several filters around the central
pass-band. This measurement is very robust, since it takes into
account the intrinsic color of each galaxy in a close region around
the spectral region of interest, rather than providing an average
continuum value in a wide wavelength range. Moreover, the continuum
determination is not affected by the emission line itself (for a
typical line such as [OII]$\lambda$3727), as would be the case for
surveys using broad-band filters.

In order to understand the scatter in Figure~\ref{fig:trumpet}, we
have analyzed the photometric uncertainties of our catalogs. The
distribution of the typical errors of our data is shown with dashed
lines. Sources whose emission in the F687W17 filter is brighter than
the average for the adjacent filters with more than the 3-$\sigma$
confidence are located below the continuous orange line. This is the
expected locus for emission-line galaxies. Note that the color to
detect a line with medium-band filters such as ours ranges between 0.1
and 0.5~mag, approximately. For filters with larger widths (e.g.,
broad-band), emission lines such as the ones we are able to detect
would be diluted by the ratio between filter widths, and thus would be
very hard to detect with broad-band data (typically having widths a
factor of 7--8 larger than the SHARDS filters, implying an effect on
the broad-band photometry of less than 0.1~mag for the brightest
lines).

In order to test how the SHARDS data perform when trying to select
ELGs, we have marked in Figure~\ref{fig:trumpet} with green symbols
the galaxies with reliable spectroscopic redshifts for which the
[OII] line would lie within the F687W17 pass-band\footnote{Note that
  these galaxies have been highlighted only because of their redshift,
  i.e., some of these galaxies may not present [OII] emission, being
  their spectroscopic redshifts based on some other spectral
  feature.}.  We are able to recover more than 90\% of the
spectroscopically confirmed galaxies with an [OII] emission-line
expected within the F687W17 filter.  A visual inspection of the
spectra for the non-selected objects (but with z$\sim$0.8) revealed
very weak or even absent [OII] emission-lines.  Therefore, we conclude
that the SHARDS data are very effective in isolating emission-line
galaxies at a similar line flux level as deep spectroscopy.

It is also interesting to note that virtually all the
spectroscopically confirmed ELGs in Figure~\ref{fig:trumpet} are
brighter than $R$$\sim$24.5. This is the spectroscopic limit of the
redshift surveys carried out in the GOODS-N field, and the typical
detection threshold for the vast majority of data taken with
state-of-the art spectrographs in 10-meter class telescopes. The
SHARDS observations reach at least 2 magnitudes fainter, and thus open
the possibility to reliably select and study fainter and/or higher
redshift ELGs.

\begin{figure}
  \begin{center}
    \hspace{-1.0cm}
    \includegraphics[angle=-90,width=9.5cm]{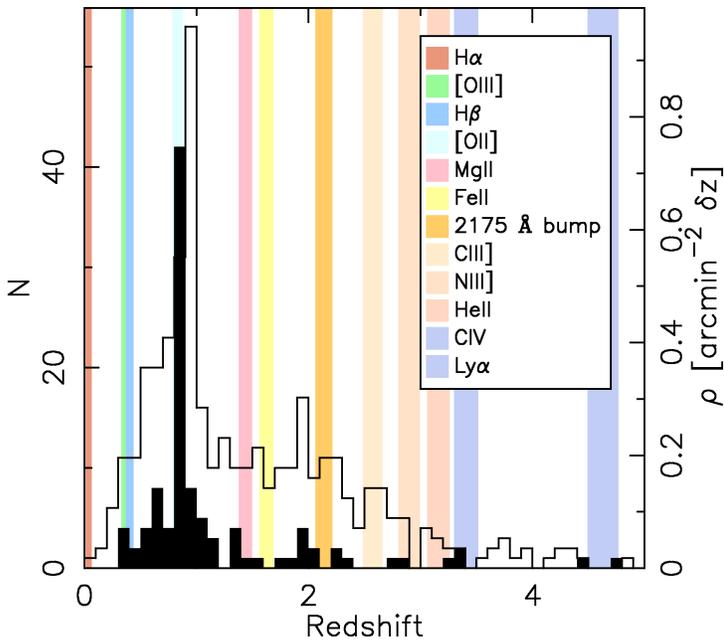}
    \figcaption{\label{fig:zdistro} Spectroscopic and photometric
      redshift distributions of the sources selected as ELGs with the
      F687W17 filter. The filled histogram shows the selected galaxies
      with spectroscopic redshifts. The open histogram shows all ELG
      candidates using photometric redshifts
      \citep{2008ApJ...675..234P}. The expected redshifts for emitters
      with some of the most typical lines (e.g., Ly$\alpha$, [OII],
      [OIII], or H$\alpha$) are marked with shadowed regions.  We also
      mark spectral features such as the Mg-Fe absorption band at
      $\sim$280~nm, or the 2175~\AA\, dust absorption bump (which
      would imprint an absorption in the galaxy SED).}
  \end{center}
\end{figure}

Figure~\ref{fig:zdistro} shows the redshift distribution of the
sources selected as ELGs by SHARDS with the F687W17 filter. The filled
histogram represents galaxies with spectroscopic redshifts, i.e.,
confirmed ELGs. The open histogram shows the photometric redshifts
\citep{2008ApJ...675..234P} for all ELG candidates. These redshift
distributions demonstrate that the ELGs selected by SHARDS are
preferentially located in the appropriate redshifts corresponding to
the most common emission lines detected in intermediate and high
redshift galaxies.  The majority of the selected sample are [OII]
emitters at z$\sim$0.8.

\begin{figure*}
  \begin{center}
    \includegraphics[angle=0,width=17cm]{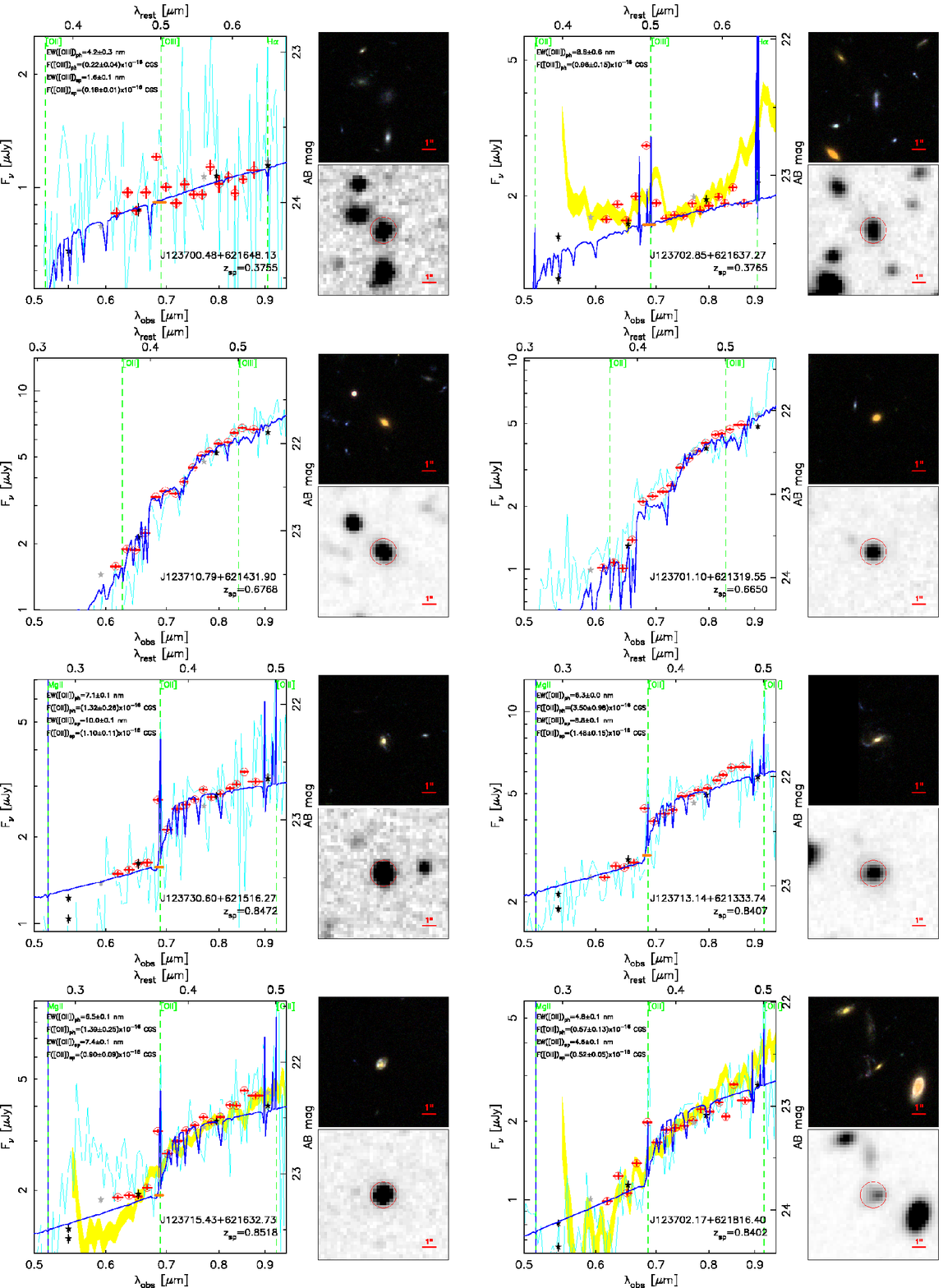}
    \figcaption{\label{fig:elgs}SEDs and postage stamps of some
      representative examples of the ELG candidates selected with the
      SHARDS F687W17 filter (from Figure~\ref{fig:trumpet}). We show
      galaxies which have already been spectroscopically confirmed.
      On the left panel for each galaxy, we depict the optical SED of
      the source. Broad-band photometric fluxes are depicted with
      black (ground-based) and gray (HST) filled stars. Red symbols
      show the SHARDS data, including uncertainties and filter widths.
      Blue lines represent stellar population synthesis models fitting
      the broad-band data, which were used to obtain an estimation of
      the photometric redshift and stellar mass
      \citep{2008ApJ...675..234P}. The cyan lines show ground-based
      spectra for each source, when available; the spectral and
      absolute flux calibration explained in
      Section~\ref{sect:reduction} have been applied. For some
      sources, we also depict with a yellow shaded area the available
      HST/ACS grism spectrum from PEARS (including uncertainties).
      The wavelength for several typical emission-lines are marked in
      green. We also give measurements of EWs and fluxes for the [OII]
      and [OIII] lines when they lie within the F687W17 filter
      pass-band, using both the spectroscopic and the photometric
      SHARDS data. On the right for each galaxy, we show postage
      stamps in the ACS bands (RGB image on top) and the SHARDS
      F687W17 filter (bottom).  North is up and East is left; image
      sizes are 10$\arcsec$$\times$10$\arcsec$, and the ELG confirmed
      candidate is marked with a red circle (radius 1\arcsec).}
  \end{center}
\end{figure*}

Our observational strategy was conceived to detect and measure
absorption bands. In this regard, absorption features such as the
Mg+Fe absorption band (at $\sim$280~nm), the Balmer and 4000~\AA\,
breaks, or the dust absorption peak at 2175~\AA\, shown in certain
extinction laws (such as the Milky Way's,
\citealt{1989ApJ...345..245C}), can mimic an emission-line in a
color-magnitude diagram such as the one depicted in
Figure~\ref{fig:trumpet} (see also Figure~\ref{fig:elgs}).

Figure~\ref{fig:elgs} shows eight examples of galaxies selected by
SHARDS as emission-line sources and counting with spectroscopic
confirmation (shown in green in Figure~\ref{fig:trumpet}). These
sources have been selected using the F687W17 filter as central band.
The two sources on the top row are examples of low redshift galaxies
detected because of their [OIII]$\lambda\lambda$4959,5007 emission
(forming the peak at z$\sim$0.3 seen in Figure~\ref{fig:zdistro} and
marked in blue and green). We compare our spectro-photometric data
with the available Keck and HST/ACS grism spectroscopy. Unlike the
noisy spectra, our data allow to study the continuum (compare with the
stellar population synthesis models shown in blue, which were fitted
to the broad-band data; see \citealt{2008ApJ...675..234P}).  Using all
the SHARDS photometric data-points and fitting them to stellar
population models, we have estimated this continuum level and,
combining this with the flux measurement for the F687W17 filter, we
have obtained line fluxes and EWs. Note that our method to estimate
the continuum is not based on the adjacent bands to the one containing
the emission line alone, but on a stellar population model fit to all
bands (except the one containing the line). When comparing to the
values obtained from the spectroscopy, we typically find smaller EWs.
We interpret this systematic difference as an aperture effect, given
that our photometric apertures are typically 1.0-1.5\arcsec\, in
radius, probably larger than the typical slit widths and extraction
sizes used in spectroscopic surveys. Note that larger apertures would
easily dilute the emission lines in a brighter continuum, thus
resulting in smaller measurements of the EW.We will present a detailed
analysis of the comparison between fluxes obtained with photometry and
spectroscopy in a forthcoming paper (Cava et al., in preparation).


On the second row of Figure~\ref{fig:elgs}, we show two examples of
galaxies located in the ELG locus in Figure~\ref{fig:trumpet}, but not
showing emission lines. These sources were selected due to the
significant flux difference between adjacent bands around the 687~nm
filter. In this case, this large color is due to the presence of the
Balmer or 4000~\AA\, break at approximately 680~nm (as clearly shown
by the available ground-based spectra and stellar population synthesis
models). These Balmer break/D(4000) sources form the peak at
z$\sim$0.7 and the redshift tail between the z$\sim$0.8 peak, which
corresponds to [OII] emitters, and z$\sim$1.2 seen in the redshift
histogram in Figure~\ref{fig:zdistro}.  The postage stamps show that
these sources are typically red spheroids, where strong 4000~\AA\,
breaks are expected. Again, our SHARDS data are able to probe this
spectral region with high accuracy and measure absorption indices such
as the D(4000). We refer the reader to Section~\ref{sect:redanddead}
for a discussion on the detection of absorption features with SHARDS
data and their use for stellar population synthesis models.

The third and fourth rows in Figure~\ref{fig:elgs} show [OII] emitters
with spectroscopic confirmation (from Keck and HST grism data)
selected in the F687W17 band ([OII] around z$=$0.8). These examples
clearly demonstrate that SHARDS data are able to detect faint
emission-line galaxies, which may be even missed by low resolution
spectroscopy such as that from HST/ACS grism data.  This is the case
for the sources in the bottom row, whose [OII] emission is not
measurable with HST/ACS data, but was confirmed with ground-based
medium-resolution spectroscopy. We would also like to highlight that
our medium-band spectro-photometric data are able to detect and
robustly measure emission-lines as faint as
$\sim$5$\times$10$^{-17}$~erg~s$^{-1}$~cm$^{-2}$.

Finally, we also point out that the postage stamps shown in
Figure~\ref{fig:elgs} demonstrate the depth and excellent image
quality of the SHARDS images, which are able to detect the majority of
the faint objects seen in the ACS data.

In summary, our science verification test has demonstrated that SHARDS
is able to detect virtually all the spectroscopically confirmed
emission-line galaxies at intermediate redshift. This opens the
possibility to extend and reach higher completeness in the samples of
ELGs in GOODS-N, and carry out a comprehensive analysis of
star-forming galaxies and AGN down to fainter magnitudes
($\sim$26.5~mag) than spectroscopic surveys. A complete analysis of
the properties of the emission-line galaxies detected by SHARDS will
be presented in a forthcoming paper (Cava et al. 2012, in
preparation).



\subsection{ELGs at very high-z}
\label{sect:laes}

\begin{figure*}
  \begin{center}
    \includegraphics[angle=0,width=16cm]{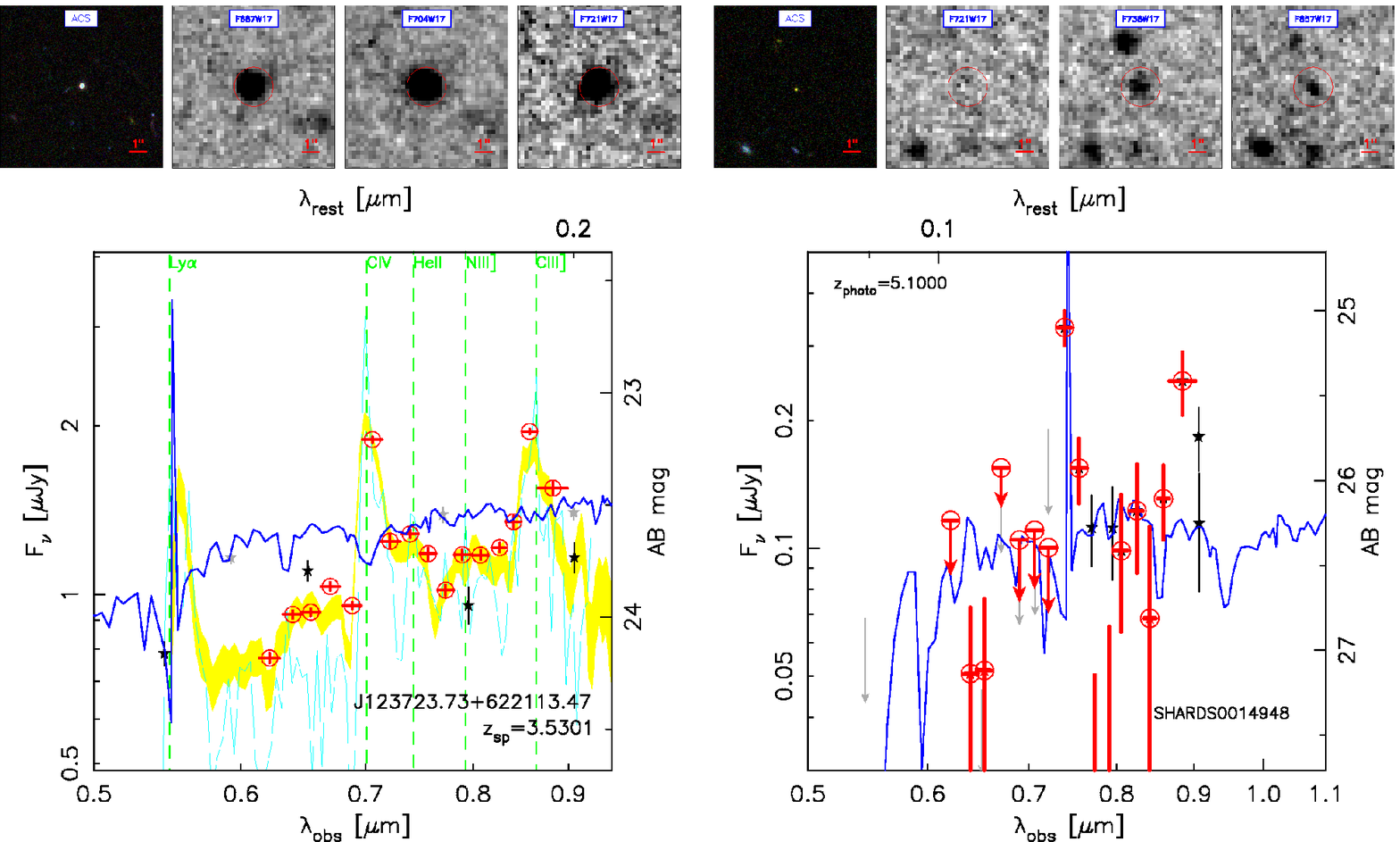}
    \figcaption{\label{fig:lae} Examples of SEDs constructed with
      SHARDS data for z$>$3 galaxies. Symbols and lines are as in
      Figure~\ref{fig:elgs}. For each source, postage stamps in
      several bands are also given. On the left, we show a
      spectroscopically confirmed AGN at z$\sim$3.5 for which the
      SHARDS spectro-photometric data provide clear detections of
      emission-lines such as CIII and CIV. On the right, we show a
      dropout source which has been selected as a candidate for a LAE
      at z$\sim$5 with a possible companion (also a LAE at the same
      redshift) to the NE \citep{2012arXiv1201.4727R}.  The SHARDS
      data (and the very few detections in broad-band imaging
      observations) are fitted with a stellar population synthesis
      model that predicts an intense Ly-$\alpha$ emission at around
      740~nm, corresponding to z$=$5.1.}
  \end{center}
\end{figure*}

The ultra-deep imaging data from SHARDS could, in principle, allow us
to detect and study sources at very high-z (z$\gtrsim$3). Indeed,
reaching magnitudes as faint as 26.5-27.0~mag in all bands from 500 to
941~nm is deep enough to detect interesting sources such as Ly$\alpha$
emitters (LAEs) at z$>3$ and up to z$\sim$6.7
\citep{2006ApJ...648....7K,
  2006ApJ...653...53B,2003ApJ...582...60O,2008ApJS..176..301O}. In
Figure~\ref{fig:lae}, we show one example of a spectroscopically
confirmed AGN at z$>$3 where SHARDS data reveal the presence of
several emission lines such as CIII]$\lambda$1909 or CIV$\lambda$1548.
The SED of this AGN shows that the spectro-photometric data from
SHARDS is completely consistent and provides a similar spectral
resolution as the HST grism data from PEARS. Our data can be used to
obtain very robust fluxes and flux densities for both emission-lines
and the continuum for a source which is as faint as the typical
spectroscopic limit of ground-based state-of-the-art spectrographs in
10m class telescopes (see the noisy Keck spectrum for this source).
Note that SHARDS data are not yet available at wavelengths below
$\sim$600~nm, but when our survey is complete we should be able to
detect and measure fluxes for Ly-$\alpha$ in this kind of object (AGN
or ELGs).

A preliminary analysis of the SHARDS data in four filters from 687 to
738~nm revealed a dozen dropout sources (in one of the four filters)
with SEDs consistent with z$>$4 LAEs and LBGs. Figure~\ref{fig:lae}
shows one of these LAE candidates at z$\sim$5, including postage
stamps and a SED. In fact, the postage stamps show two LAE
candidates, both appearing in the F738W17 filter (one of them being
almost undetected in the ACS bands). The galaxy shown in the SED is a
dropout in the F721W17 filter. Its emission is booming in the F738W17
filter, and then becomes fainter again (but detected) in redder
filters. This points out to a Lyman break located within the two
mentioned SHARDS filters, which would imply a redshift around z$=$5.
This figure demonstrates the potential of the SHARDS ultra-deep
medium-band survey to select magnitude 26--27 emission-line galaxies
at very high redshifts.  We refer the reader to Rodriguez-Espinosa et
al.  (2012b, in preparation; see \citealt{2012arXiv1201.4727R}) for a
more detailed discussion on the detection and properties of the LAEs
detected by SHARDS.


\section{SHARDS Science verification: absorption systems}
\label{sect:redanddead}

The main goal of the SHARDS project is to analyze in detail the
stellar populations in massive galaxies at high redshift, especially
those that are already evolving passively. The observational strategy
of the survey was devised to be able to construct rest-frame
UV/optical SEDs for this kind of sources with enough spectral
resolution to measure absorption indices correlated with important
physical parameters such as the age of the stellar populations.

The main absorption index targeted by SHARDS for high-z passively
evolving sources is the Mg index, \Mg. This index probes several
absorption lines (e.g., MgI$\lambda$2852,
MgII$\lambda\lambda$2796,2804, FeII$\lambda\lambda$2600,2606) and has
been shown to be an extremely reliable means to detect quiescent
massive galaxies. In addition, the absorption lines can be used to
easily distinguish the SED of a quiescent massive galaxy from the
featureless spectrum of a dusty starburst \citep{2005ApJ...631L..13D}.
The relative intensity of these absorptions can be measured with the
\Mg\, spectral index, directly linked to the age of the stellar
population \citep[e.g.,][]{1993ApJ...405..538B}. The index is easily
and robustly measurable even in low resolution spectra (R$<$100,
\citealt{2005ApJ...631L..13D}). The \Mg\, index has been successfully
used in the past to obtain redshifts and ages of stellar populations
in massive galaxies at high-z
\citep{1997ApJ...484..581S,2004Natur.430..184C,2004ApJ...614L...9M,
  2005MNRAS.357L..40S,2005ApJ...631L..13D,2004MNRAS.350.1322F,
2012AJ....144...47F}.

Other interesting absorption features probed by SHARDS (for galaxies
at different redshifts) are the Balmer and 4000~\AA\, break, or the Ca
HK lines, the G-band, and the Mg$_1$, Mg$_2$ and TiO$_2$ molecular
bands (among others). All these have been extensively used to study
stellar population ages \citep[e.g.,][see also the review by
\citealt{2006ARA&A..44..141R}]{1984ApJ...287..586B,1994ApJS...95..107W,
  1994ApJS...94..687W, 1999ApJ...527...54B,2000ApJ...532..830L,
  2003ApJ...587L..79F,2003MNRAS.341...54K,2004ApJ...616...40F,
  2005MNRAS.362...41G,2006ApJ...649L..71K,2011ApJ...743..168K,
  2012ApJ...746..188M}.  The spectral resolution of the SHARDS dataset
is adequate for measuring the Balmer or D(4000) breaks with high
accuracy. For galaxies with spectroscopic redshifts or very accurate
photometric redshift (errors below 1\%), we have very similar
accuracies to those achieved in standard spectroscopic studies of
nearby galaxies and synthesis models, which typically use bands of
10--20~nm around the feature \citep{1983ApJ...273..105B}. In the case
of the molecular bands, the widths of the SHARDS filters are larger
than the amplitude of these absorptions, but, given their strength,
they should have a measurable effect on the photometry through filters
such as ours. In any case, our filters are narrower (by at least a
factor of $\times$2) than those used by other optical and NIR
medium-band surveys (such as MUSYC or NMBS). This allows us to probe
these features with higher resolution.  In addition, as demonstrated
in the previous Section, SHARDS data are also very sensitive to
emission lines, so they can be used to detect low level residual
(unobscured) star formation in massive galaxies at high-z and confirm
their quiescent state (jointly with other methods, such as the study
of the MIR/FIR emission).

\subsection{Study of quiescent massive galaxies at high-z}

In this Section, we demonstrate the power of the SHARDS dataset to
accurately measure absorption indices and study high-z massive
galaxies. To do so, we focus on quiescent massive galaxies with
reliable spectroscopic redshifts above z$=$1.  Rather than measuring
and studying one specific absorption index, we combine all the
photometric information for each source and fit the entire SED with
stellar population synthesis models, i.e., we include SHARDS fluxes
and also broad-band data up to the MIR wavelengths probed by {\it
  Spitzer}/IRAC. We, therefore, use all the (photometric) observations
available for our sample of quiescent massive galaxies to carry out
the most reliable analysis possible of their stellar population
properties. We concentrate our study on z$\sim$1-2, a critical
redshift range in which massive quiescent galaxies assembled a
significant fraction of their mass
\citep{2007A&A...476..137A,2008ApJ...675..234P,2010ApJ...709..644I,
  2011ApJ...727...51N,2011ApJ...739...24B}. In this redshift interval,
the D(4000) and/or \Mg\, absorption features are probed by the SHARDS
data.

The sources compiled for this science demonstration exercise have been
extracted from several sources in the literature. The sample includes:

\begin{itemize}
\item the three galaxies classified as {\it red nuggets} whose
  dynamical masses were studied in \citet{2010ApJ...717L.103N} and are
  covered by our data. All the sources in this paper within our
  surveyed area were detected in all SHARDS individual filters, but
  some of the other sources from Newman et al.'s sources lie outside
  of our FOV.
\item the four galaxies at z$>$1 from \citet{2009ApJ...706..158F}, who
  concentrated their study of early type galaxies on intermediate
  redshift sources, but had a few z$>$1 galaxies in their sample.
  Given that the galaxies in this paper count with
  pseudo-spectroscopic redshifts from HST optical grism data, we
  compared the quoted redshifts with our photometric redshifts
  \citep{2008ApJ...675..234P} and only selected the four galaxies with
  consistent estimations.
\item Finally, we also selected those galaxies with spectroscopic
  redshifts that, according to their observed optical and NIR colors,
  qualify as Extremely Red Objects (ERO, following the definition in
  \citealt{1988ApJ...331L..77E} and \citealt{2000AJ....120..575Y}),
  and/or Distant Red Galaxies (DRGs) at z$>$1
  \citet{2003ApJ...587L..79F}.
\end{itemize}

In order to avoid dusty starbursts in this study and concentrate only
on massive galaxies that are evolving passively, we only considered
galaxies undetected in the MIPS 24~\mic\, data, which reach fluxes as
low as 30~$\mu$Jy in GOODS-N (reduction and catalogs described in
\citealt{2005ApJ...630...82P} and \citealt{2008ApJ...675..234P}). This
sets an upper limit of 1--10~M$_\sun$~yr$^{-1}$ at z$=$1.0--1.5.  The
final sample considered in the following discussion is composed of 27
galaxies at 1.01$<$z$<$1.43, with virtually all of them having masses
above 10$^{10.5}$~M$_\sun$.  Their main properties are given in
Table~\ref{table:massive}. Note that we have only considered galaxies
with a spectroscopic redshift, so the sample is highly biased towards
low redshift (close to unity, the limit lower of our sample, 75\% of
the sample is at z$<$1.3 and the average redshift is
$\bar{\mathrm{z}}$$=$1.17) and not (necessarily) representative of the
entire population of massive quiescent galaxies at z$\sim$1.

\input{tab2_2c}

\subsection{Estimating stellar population properties from SED
  synthesis models}

Our goal in this Section is to improve the characterization of the
stellar populations of quiescent galaxies by using the unprecedented
spectral resolution and depth provided by the SHARDS photometric
observations.  Recent results based on data with similar medium-band
filters and (low resolution) grism spectroscopy have already shown the
power of probing key spectral features on the SEDs of quiescent
galaxies, such as continuum breaks and absorption lines/bands, to
place more robust constraints on relevant properties of these galaxies
such as the age, stellar mass, or star-formation history
\citep{2009ApJ...706..158F,
  2010ApJ...719.1715W,2011ApJ...735...86W,2010ApJ...718L..73V}.

Here we study the stellar properties of a sample of 27 massive
quiescent galaxies with spectroscopic redshifts above z$=$1 by fitting
their UV-to-MIR SEDs with stellar population synthesis (SPS) models.
To calibrate the impact of using the SHARDS data in this exercise, we
conduct two kinds of tests. First, we show how including the SHARDS
medium-band data allows us to decrease those uncertainties in the
derived stellar parameters linked to well-known degeneracies (such as
age-dust, or age-SFH). To do so, we analyze the connection between the
photometric uncertainties and the best-fit solutions in the parameter
space formed by the relevant stellar properties.  Second, we study the
different results obtained by fitting the SEDs with different sets of
SPS models extracted from several of the most common libraries found
in the literature.

We start the analysis of the SEDs of massive quiescent galaxies at
z$\sim$1 by describing the set of libraries, the modeling assumptions,
the parameter space, and the statistical approach to the analysis of
the different solutions and degeneracies.

The UV-to-MIR SEDs for the sample of 27 massive passively evolving
galaxies at z$>$1 were fitted with SPS models from several of the most
common libraries found in the literature. We considered star formation
histories (SFHs) following general exponential decaying models
($\tau$--models) using the predictions from the following codes:

\begin{itemize}
\item the \citet{2003MNRAS.344.1000B} models (BC03) with a
  \citet{2003ApJ...586L.133C} IMF. This was chosen as our fiducial
  model (BC03-chab, hereafter).  In this section and the following, we
  discuss the properties of our galaxies as inferred by other models
  in comparison with those obtained with this library and IMF.
\item the Charlot \& Bruzual (2009) models (CB09) with a Kroupa and
  Chabrier IMFs (CB09-krou and CB09-chab, hereafter). These models are
  thought to be an improvement over the BC03 library with a better
  treatment of some evolutionary stages of stars.
\item the \citet{2005MNRAS.362..799M} models (M05) with a Kroupa IMF
  (M05-krou). These models should also give us information about the
  effects on the derived parameters of specific stellar evolutionary
  stages such as the thermally-pulsating AGB phase.
\item the PEGASE code \citep[P01]{1997A&A...326..950F}, assuming a
  \citet{2001MNRAS.322..231K} initial mass function (IMF). These are
  referred to as P01-krou models.
\end{itemize}

In all cases, we considered the following mass limits for the IMF:
0.1$<$$\mathcal{M}$$<$100~$\mathcal{M}_\sun$.

The fitting procedure is explained in detail in
\citet{2003MNRAS.338..525P} and \citet{2008ApJ...675..234P}. Briefly,
the photometry is compared with the models assuming an exponentially
decreasing SFH characterized by a $\tau$ parameter which runs from
1~Myr (almost identical to a single stellar population, SSP) to
100~Gyr (almost constant SFH). We run models corresponding to the
different discrete values of the metallicity given for each library,
which run from $Z_\sun$/200 to 2.5$Z_\sun$ for BC03 and CB09 models,
$Z_\sun$/200 to 5$Z_\sun$ for P01, and $Z_\sun$/50 to 2$Z_\sun$ for
M05 models. We considered an extinction parametrized with a $V$-band
attenuation, A(V), with values ranging from 0 to 2~mag, and assumed
the extinction law of \citet{2000ApJ...533..682C}.  Finally, we probed
all ages between 1~Myr and the age of the Universe corresponding to
the source redshift (up to $\sim$6~Gyr for the closest galaxy) to
search for the best fitting model minimizing a reduced $\chi^2$
maximum-likelihood estimator.

In order to study the uncertainties in the derived parameters and to
take into account the possible degeneracies in the solutions, we run
Montecarlo simulations for each galaxy. The method consisted in
randomly varying the photometric data points with a Gaussian
distribution of width equal to the photometric uncertainty, and
repeating the fit again with all possible models. We run the code 1000
times and then analyzed the set of solutions. This analysis identified
the clusters of solutions with a k-means method.  Each cluster was
assigned with a statistical significance, given by the fraction of the
1000 different solutions belonging to the cluster, i.e., the relative
number of solutions which provide similar results and are grouped as a
single solution identified by a median value and a scatter in the
multi-dimensional space formed by the fitted parameters. Note that we
did not use the reduced $\chi^2$ values to assign statistical weights
to each cluster of solutions, since $\chi^2$ differences between them
are not statistically relevant. Typically, 1 to 4 clusters of
solutions were identified for each galaxy above a statistical
significance of 10\%.  In virtually all cases (all galaxies and
models), one solution was dominant with more than 50\% of the solution
data points belonging to its cluster. Note that different clusters may
have different typical reduced $\chi^2$ values, i.e., the data can be
better reproduced by some combination of parameters corresponding to
one cluster. The different solutions provided information about the
typical degeneracies of the study of stellar populations, such as the
age-metallicity degeneracy or the one linked to $\tau$ (SFH) and age.
Table~\ref{table:massive} presents the different solutions (the
statistically most significant and the secondary solutions) for each
galaxy and the statistical weight of each of them.

\subsection{Evaluation of the stellar population synthesis modeling
  with SHARDS data}

\begin{figure*}
  \begin{center}
    \hspace{-1.0cm}
    \includegraphics[angle=-90,width=6.1cm]{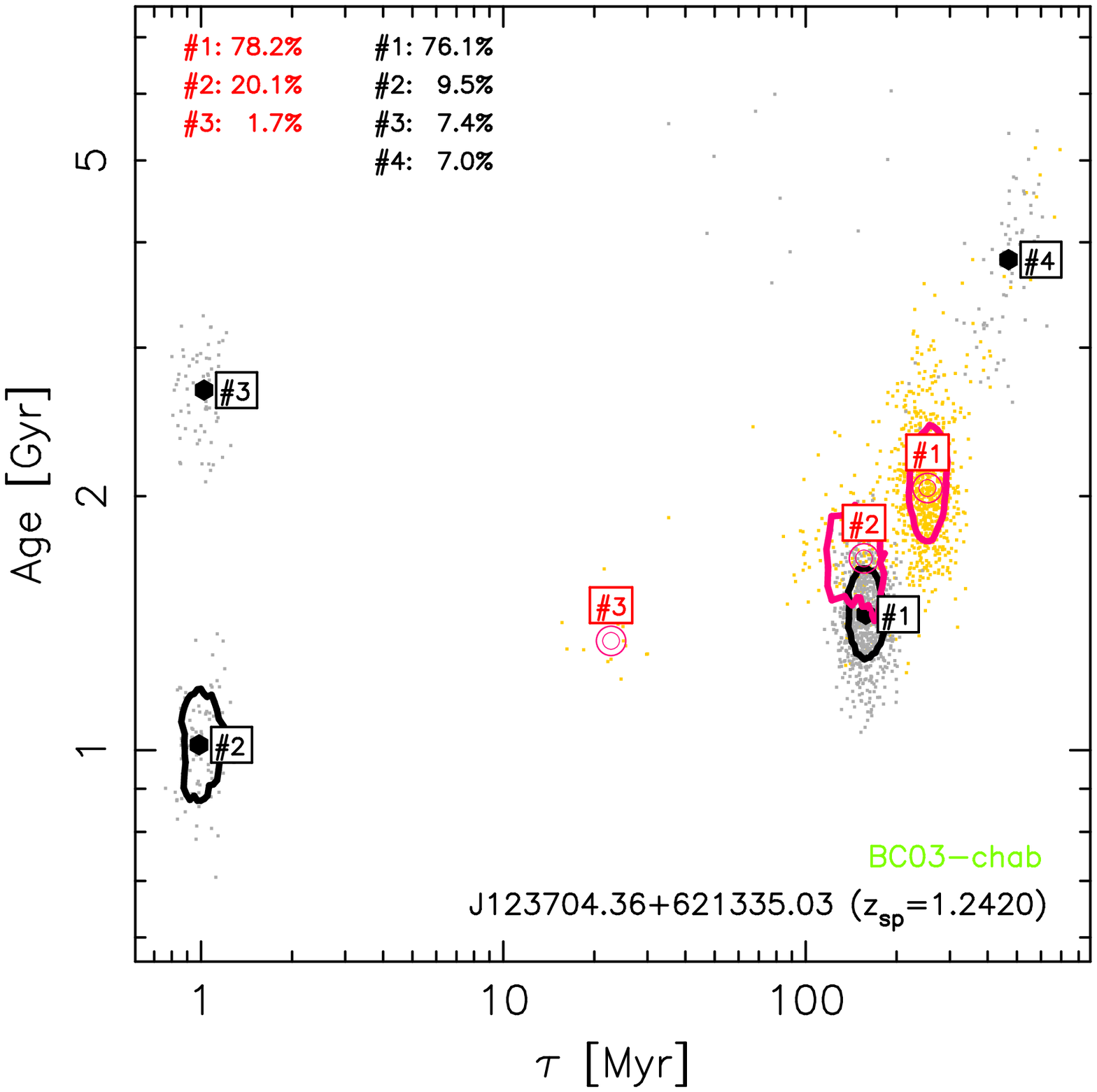}
    \includegraphics[angle=-90,width=6.1cm]{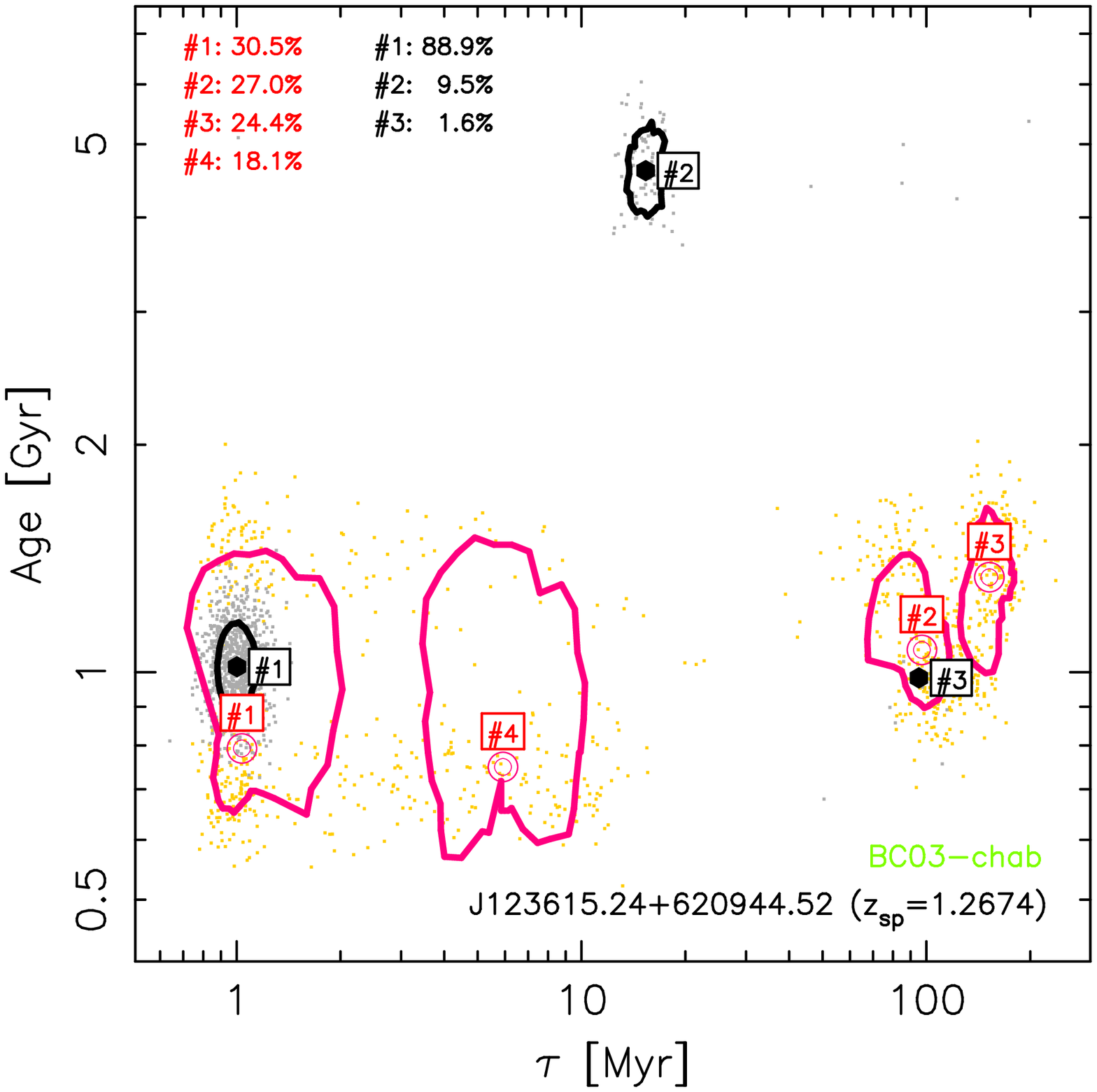}
    \includegraphics[angle=-90,width=6.1cm]{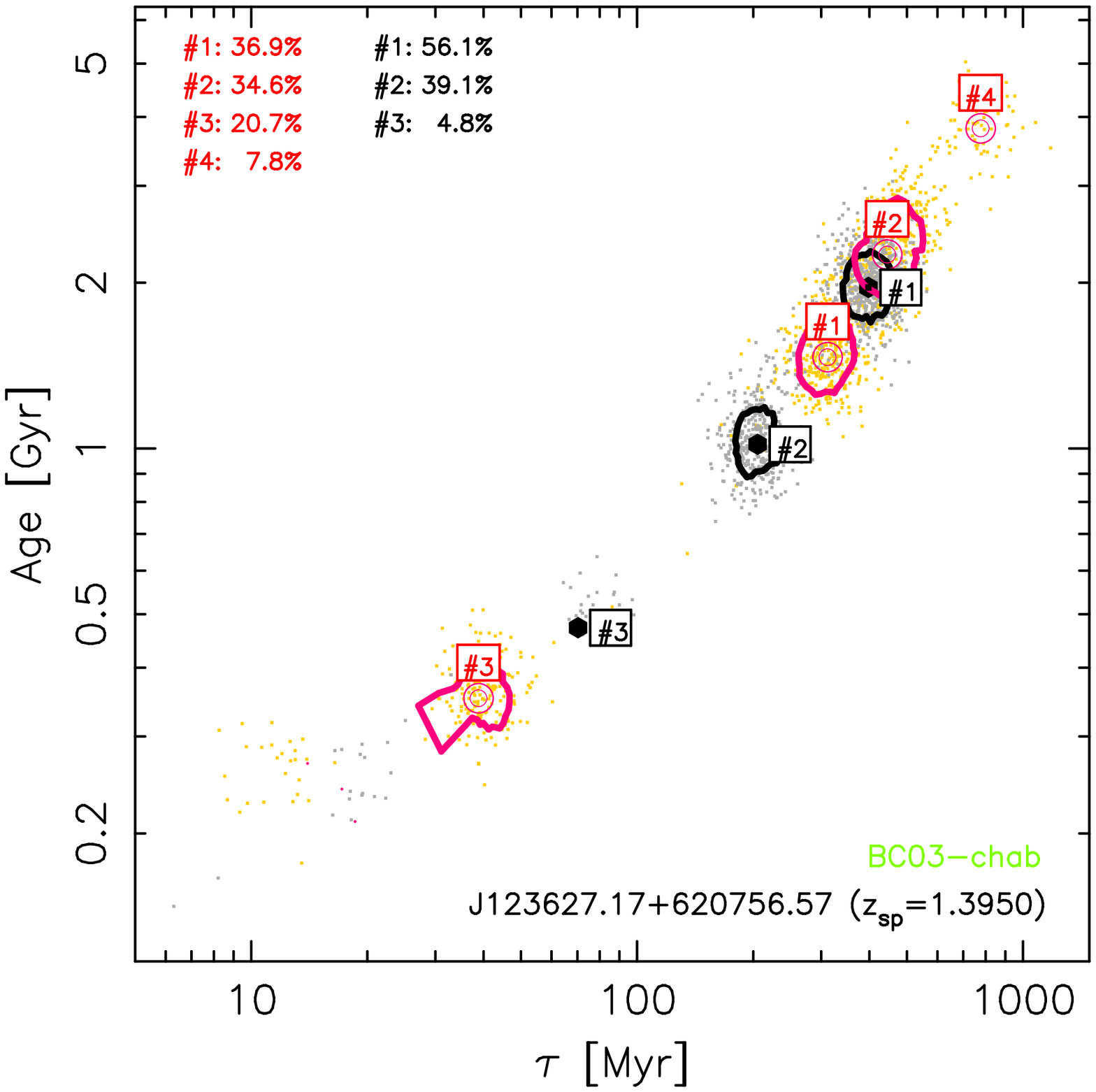}
    \figcaption{\label{fig:shards_vs_bb} Age and e-folding time
      results for the stellar population synthesis modeling (using our
      fiducial library, BC03-chab) of the SEDs for the three massive
      quiescent galaxies at z$>$1 presented in Figure~\ref{fig:drg}.
      For each galaxy, orange points show the solutions when fitting
      UV-to-MIR broad-band data alone, while gray points show the
      results when adding the medium-band SHARDS data also. Average
      values for each cluster of solutions are marked with open
      symbols for the results using broad-band data alone, and filled
      circles for the solutions also using SHARDS data. Contours show
      2$\sigma$ uncertainties: red for fits to only broad-band, black
      for fits adding SHARDS fluxes.  Statistical significances of
      each cluster of solutions are also given. Note that the scales
      are adjusted to each galaxy, but our modelling procedure has
      probed e-folding times from 1~Myr to 100~Gyr, so only a small
      part of the considered parameter space is shown.}
  \end{center}
\end{figure*}

In order to test how the spectro-photometric data from SHARDS helps in
the SPS modeling of high-z galaxies, we compare in
Figure~\ref{fig:shards_vs_bb} the results obtained when fitting only
the broad-band data with those obtained when adding the SHARDS data
for three representative galaxies of our sample of quiescent massive
galaxies at z$>$1. The details about the fits and results of the SPS
modeling for these galaxies are presented in Figure~\ref{fig:drg} and
Section~\ref{sect:example}.

For the first galaxy, J123704.36$+$621335.03 (z$=$1.2420), the fits to
the broad-band photometry alone provide a dominant solution
characterized by an e-folding time $\tau$$\sim$200~Myr and an age of
around 2~Gyr. There is another less statistically significant solution
involving a shorter and younger burst ($\tau$$\sim$150~Myr and age
1.8~Gyr), and a negligible fraction of solutions would favor even
lower e-folding times and ages.  When adding the SHARDS data to the
fits, a solution with $\tau$$\sim$150~Myr and t$\sim$1.5~Gyr shows the
largest statistical weight.  This solution is consistent, within
uncertainties, with one of the solutions found when fitting the
broad-band data alone, although a slightly younger burst is favored in
the SHARDS fluxes (age of $\sim$1.5~Gyr instead of $\sim$2~Gyr).

In the case of J123615.24$+$620944.52 (z$=$1.2674), the broad-band
data provide a wide range of roughly equally significant solutions.
Each one has 20--30\% statistical significance. They show different
values of extinction (differences of up to 0.5~mag from one solution
to another), metallicity and e-folding times (from an SSP to almost
200~Myr), while ages do not vary wildly (all solutions predict ages
slightly above 1~Gyr). The SHARDS data break most of this degeneracy
and majorly favor a 1~Gyr old SSP with A(V)$=$1~mag.  This is one of
the largest extinctions in our sample, which could explain the
relatively high flux at 8~$\mu$m, maybe revealing the presence of PAHs
(see Figure~\ref{fig:drg}).  Moreover, this galaxy has a very red
companion located 3\arcsec\, away to the NE (26 kpc), also at
z$_\mathrm{sp}$$=$1.2630. This source is clearly detected at 24\mic,
which might be hiding some residual MIPS emission for our galaxy, and
provides an SFR$\sim$20~M$_\sun$~yr$^{-1}$. A possible interaction
between the two objects could have switched on the star formation and
then stopped recently, thus explaining the relatively high extinction.

In the case of J123627.17$+$620756.57 (z$=$1.3950), both the
broad-band data alone and the whole photometric information including
SHARDS data favor at the approximately 60\% significance level a SFH
with an e-folding time around 400~Myr and $\sim$2~Gyr old, with
A(V)$=$0.2~mag. The SHARDS data help to discard some solutions found
with only broad-band data which indicate very young and short bursts
with high extinction (1~mag).

\subsection{Examples of the stellar population synthesis results
  including SHARDS data}
\label{sect:example}

\begin{figure*}
  \begin{center}
    \includegraphics[angle=0,width=14cm]{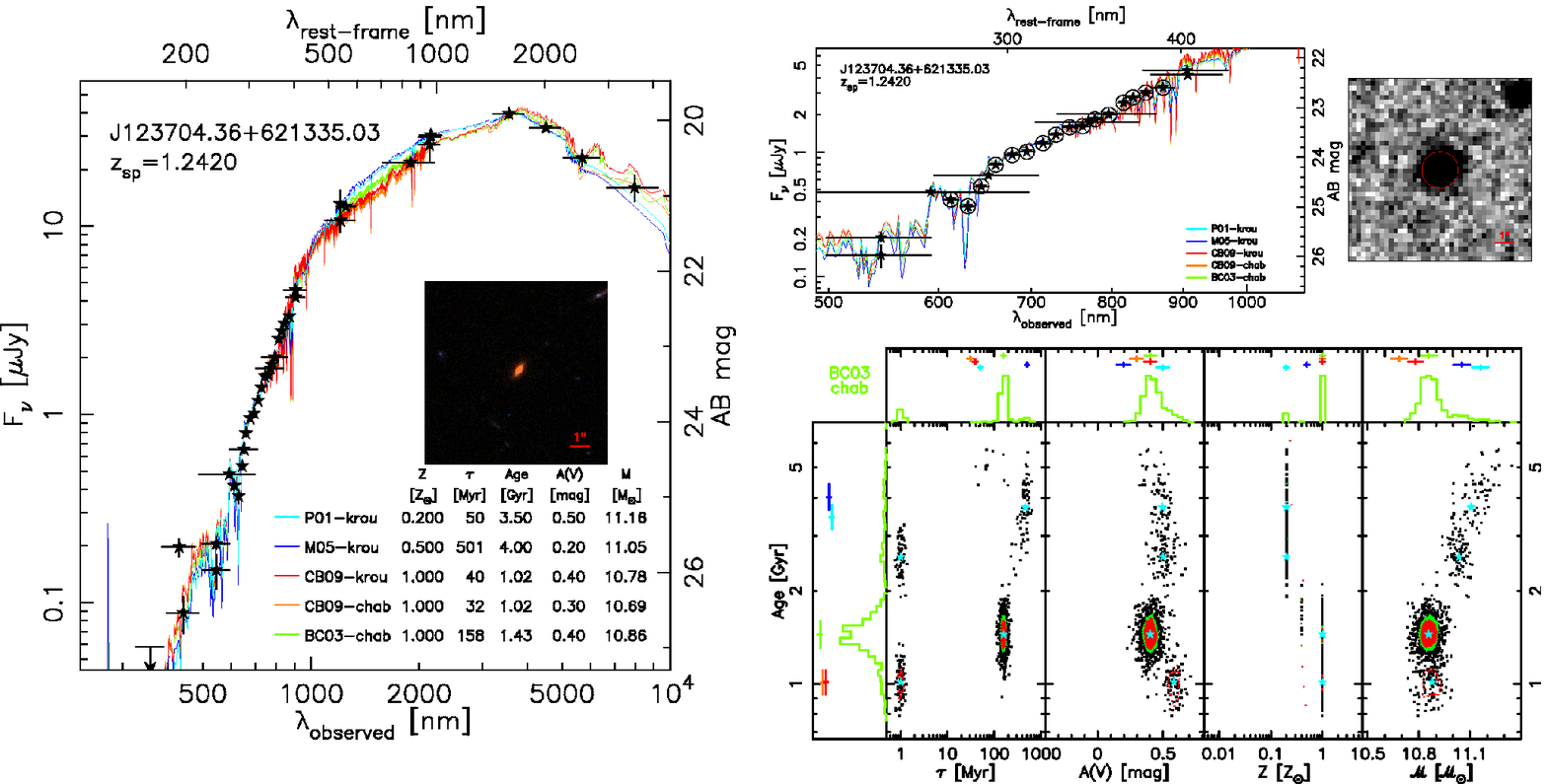}
    \includegraphics[angle=0,width=14cm]{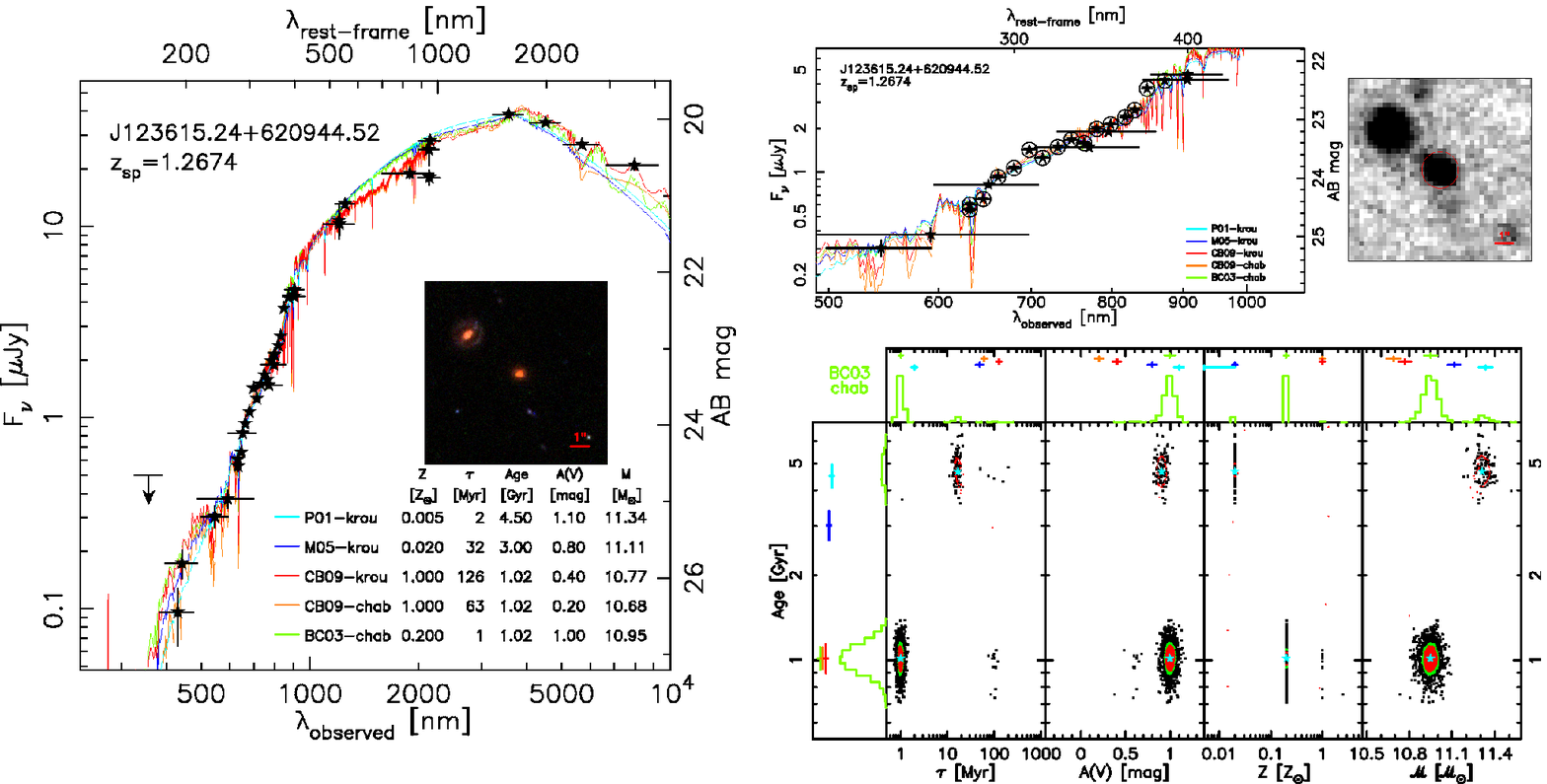}
    \includegraphics[angle=0,width=14cm]{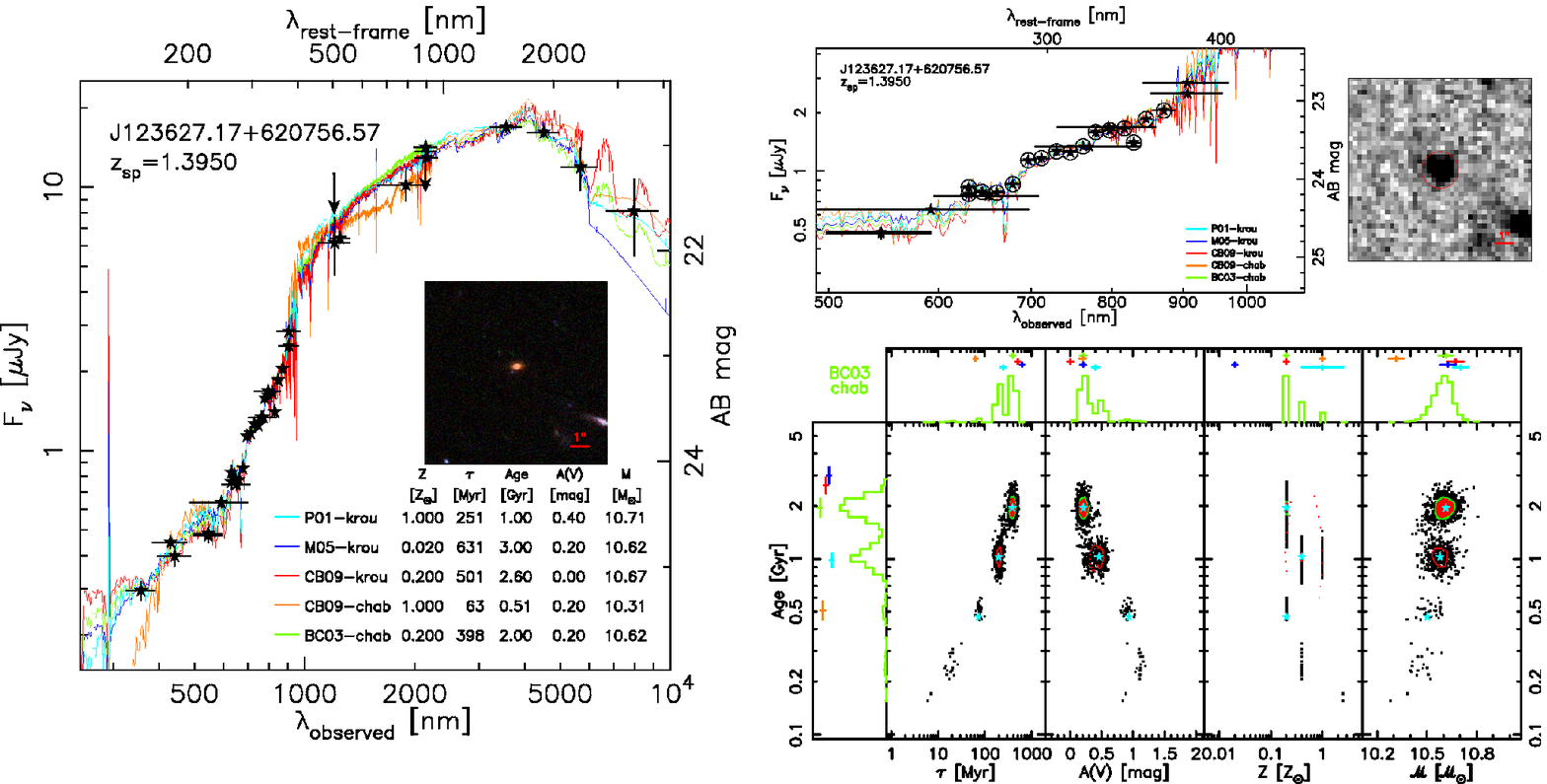}
    \figcaption{\label{fig:drg} Three examples (one per row) of the
      stellar population synthesis models carried out on quiescent
      massive galaxies with spectroscopic redshifts at z$>$1. Each row
      shows the entire UV-to-MIR SED of a galaxy, jointly with the
      best fits for different modeling libraries (see text for
      details), and an RGB 10\arcsec$\times$ 10\arcsec postage stamp
      of the source built from HST data. The results for the best
      fitting model are also given for each library, including
      ($Z/Z_\sun$,$\tau/Gyr$,t/$Gyr$,A(V),M/M$_\sun$).  Photometric
      data points include uncertainties and the width of the filter.
      To the right of the general SED, we show a close-up of the
      spectral region covered by SHARDS. On the top-right corner, we
      depict a postage stamp of the galaxy in the F687W17 SHARDS band.
      Its size is 10\arcsec$\times$ 10\arcsec, and the 1\arcsec radius
      red circle marks the galaxy.  The bottom-right panel shows the
      position of the different model solutions derived with our
      Montecarlo method to determine uncertainties and degeneracies.
      The data points correspond to our fiducial model: BC03--chab. We
      depict all the 1000 solutions in different planes: age-$\tau$,
      age-extinction, age-metallicity, and age-mass. For the most
      statistically significant cluster, we mark the average solution
      with a star, the 1$\sigma$ area is filled in red, and the
      2$\sigma$ region in green. For the rest of solutions with a
      significance above 10\%, we mark the average with a star, and
      the area corresponding to 1$\sigma$ with a red line. Histograms
      for the different parameters are also given, marking the average
      and 1$\sigma$ values for the different models mentioned in the
      text (color code as in the left panel).}
  \end{center}
\end{figure*}

In Figure~\ref{fig:drg}, we show the complete SEDs of the examples
presented in the last subsection. We also present the best fitting
models for the different libraries considered in this paper, jointly
with a close-up look into the rest-frame UV/optical range probed by
the SHARDS data. We include several plots showing the results of the
synthesis models and the clustering analysis of the solutions.
Finally, we also provide postage stamps in the HST and SHARDS filters.

The galaxy shown at the top of Figure~\ref{fig:drg} lies at
z$\sim$1.2420 and our currently available SHARDS data do cover the
\Mg\, absorption feature. The SPS analysis for the BC03-chab fiducial
models shows that there are at least 4 different clusters of
solutions. The dominant cluster (76\% statistical weight) is
consistent with a short starburst ($\tau$$\sim$150~Myr) with solar
metallicity and relatively low extinction, A(V)$=$0.4~mag. Among the
other clusters, presenting all of them statistical significances below
10\%, we find a longer, lower abundance, and older burst
($\tau$$\sim$600~Myr, t$\sim$4~Gyr, $Z$$=$$Z_\sun$/5), and very short
bursts (almost SSPs) with an age between 1 and 3~Gyr. These clusters
share very similar extinction values [A(V)$=$0.4~mag], with shorter
bursts being more attenuated. Note that the latter are significantly
less statistically representative than the former (i.e., models do
favor a 150~Myr burst).  We estimate a stellar mass within an interval
around 10$^{10.8-11.1}$~M$_\sun$.  Interestingly, although all
different libraries obtain similarly good fits (i.e., similar reduced
$\chi^2$ values), they provide significantly different values for the
relevant parameters, especially for the age and $\tau$ values.
P01-krou and M05-krou favor the longest, oldest, less metallic, and
higher mass solution, while BC03 and CB09 models for different IMFs
are all consistent with shorter and younger bursts.

The second example in Figure~\ref{fig:drg} shows a z$=$1.2674 galaxy
where we have also been able to measure the \Mg\, absorption (but only
the reddest part) with the SHARDS data available so far, apart from
other spectral features such as the bluest part of the 4000~\AA\, (or
Balmer) break. The SPS degeneracies are considerably smaller than in
the first example.  Indeed, there is a dominant solution corresponding
to a star-forming burst with $\tau$$\sim$1~ Myr (a SSP), age around
1~Gyr, A(V)$\sim$1.0~mag, sub-solar metallicity, and mass around
10$^{11.0}$~M$_\sun$.  Although this solution clearly dominates (90\%
probability) for the fiducial model, BC03-chab, other libraries again
achieve different results, although with very similar statistical
weight (i.e., one solution dominates for all different models). Note
that the different libraries provide very similar results in the
UV/optical part of the SED, but differ significantly in the $H$-band
and beyond the 1.6\mic\, bump, where BC03 and CB09 models reproduce
the data more closely.

Finally, the last example in Figure~\ref{fig:drg} shows one of the
sources at the highest redshifts in our sample of massive quiescent
galaxies, z$\sim$1.4. In this case, two solutions show similar
significances, with e-folding times $\tau$$\sim$200-400~ Myr, age
around 1--2~Gyr, A(V)$=$0.2--0.4~mag, sub-solar metallicity and mass
10$^{10.6}$~M$_\sun$. Similarly to the behavior of the different
libraries in the previous example, in this case different models also
obtain different estimations for the relevant parameters, especially
the age. $H$-band data probing the wavelength range around the
rest-frame $RI$-bands would most probably solve completely the
degeneracy between the models.



\subsection{Goodness of the fits}

It is interesting to discuss the properties of the sample and the
goodness of the fits for the different SPS libraries. Concerning the
latter, a very significant fraction of the whole sample, around 95\%
of all galaxies, is better fitted by the BC03-chab models, i.e., these
models provide better reduced $\chi^2$ values than any other code and
IMF.  This is the main reason for our choice of the BC03-chab models
as fiducial in this paper. The rest of best fits correspond to the
CB09-krou and P01-krou models ($\sim$2--4\% of sources each).  After
the best fitting code, the second most preferred library and IMF is
M05-krou, which gives the second best solution for 93\% of the
sources, with the rest of second best solutions corresponding to
P01-krou. Note that among the two models sharing the same library but
with different IMFs, CB09-chab and CB09-krou, the Kroupa IMF provides
the best results, but since the best fitting code is BC03-chab, it
seems that the goodness of the fit is more related to the SPS library
than to the choice of IMF, i.e., based on the analysis of the goodness
of the fits we cannot claim that one IMF works better than the other.

To further test this result indicating a preference for the BC03
models for the vast majority of sources, we repeated the SPS analysis
but only using the broad-band data, i.e., without introducing the
SHARDS fluxes. Interestingly, the best fits were then achieved by the
BC03 models for 15\% of the sample, while CB09 and M05 models obtained
the best fits for 20--30\% of the galaxies, and P01 for just 8\%. This
clearly demonstrates than the UV/optical part (more precisely, the
200$\lesssim$$\lambda_\mathrm{r-f}$$\lesssim$400 nm interval) of the
SED is better reproduced by the BC03 models, while the fits to the
global broad-band photometry up to the IRAC bands is similarly
reproduced (roughly) by all codes (except P01).


Concerning the statistical significance of the best solution, typical
values for all models are always higher than 75\%, although they could
be as low as 40\% in a few cases.  Three quarters of the sample have a
dominant solution above 50\% for all models and IMFs. The actual
individual values of the statistical significance of the different
clusters of solutions are similar for the different codes, but for
virtually all sources the most concentrated solution is given by the
BC03-chab models (the typical probability for the best solution is
90\%) and M05-krou (88\%). For the fits to broad-band fluxes only,
these figures are smaller by $\sim$20\%.
Figure~\ref{fig:shards_vs_bb} showed the comparison of SPS solutions
obtained when using only broad-band data, and when adding the SHARDS
medium-band photometry, for the sources plotted in
Figure~\ref{fig:drg}.


\subsection{Statistical properties of the sample and systematics}


Statistics about the main properties of the stellar populations in our
sample of massive quiescent galaxies are given in
Table~\ref{table:stats}. This statistical information is presented for
the distinct SPS codes, and discussed in this section.

According to our fiducial BC03-chab models, the median stellar mass of
our sample of 27 massive quiescent galaxies at z$>$1 is
10$^{10.7}$~M$_\sun$. Typically, their stellar population is 1.5~Gyr
old and was formed in a burst with solar or slightly sub-solar
metallicity with a SFH characterized by an e-folding time
$\tau$$\sim$150~Myr, and currently presents a moderate extinction
A(V)$\sim$0.5~mag.

If we consider the average of the different stellar mass estimates
obtained with the distinct codes and IMFs for a given galaxy, its
typical rms is 0.15~dex, and for some galaxies the rms can be as high
as 0.27~dex.  Most of this rms is due to systematic differences
between the results obtained with different models and IMFs. According
to our fiducial models, BC03-chab, masses for individual sources run
from 10$^{10.3}$~M$_\sun$ to 10$^{11.2}$~M$_\sun$. BC03-chab provide,
on average, larger masses than CB09-krou by $\sim$20\% when comparing
values for each galaxy, with differences as large as a factor of 2 (in
both directions). CB09-chab models give masses smaller than BC03-chab
by 30\% on average, and up to a factor of 3. We note that differences
between the two CB09 models with different IMFs are not just a
constant offset, but are variable from galaxy to galaxy and dependent
on other parameters such as age.  M05-krou and P01-krou provide
heavier masses than BC03-chab by 15\% and 40\% on average,
respectively, with differences as high as a factor of 3--4. Note that
the effects of the TP-AGB phase, which could in principle be studied
comparing classical models with libraries such as M05 or CB09, are
complex, sometimes producing larger and sometimes smaller masses.

Extinctions show better agreement between different models. The
average value for the sample is A(V)$=$0.5~mag for our fiducial
models, with a typical uncertainty of 0.2~mag. This is also the
typical average systematic offset from one model to another. All
models provide systematically lower extinctions than the BC03-chab
models.

Our sample presents ages in the range 0.5 to 3.5~Gyr for our fiducial
model BC03-chab. On average, BC03 and CB09 models provide similar ages
within 10\%, but differences from model to model among these libraries
range from 0.2 to 3~Gyr for individual galaxies. M05-krou and P01-krou
models provide, on average, older ages by 90\% and 30\%, respectively.
These ages and systematic differences are in good agreement with the
results found in the literature for z$\sim$1 galaxies
\citep[see, e.g.][]{2009ApJ...690.1866H,2010ApJ...719.1715W,2012ApJ...745..179W}.


The typical star formation time-scale for our sample is
$\tau$$\sim$150~Myr for our fiducial models.  For CB09 models and the
two IMFs considered, $\tau$ values are, on average, 25\% shorter.  M05
and P01 results indicate longer e-folding time-scales by 70\% and
20\%, respectively.

Concerning metallicities, for the BC03 models the average value is
solar, and that is also the case for all the other codes, except M05
models, which provide an average value of 0.4$Z_\sun$. Note that a
photometric study such as ours has severe difficulties in determining
accurate stellar metallicities and breaking degeneracies involving
age, extinction and metallicity
\citep{1986stpo.meet..167O,1994ApJS...95..107W,2009ApJ...696..348W},
but it is significant that all models provide consistent values near
or slightly below the solar value (consistently with other works such
as \citealt{2009ApJ...690.1866H} or \citealt{2009ApJ...706..158F}; see
also \citealt{2003ApJ...591..878D} and \citealt{2007MNRAS.381L..74K}).

\input{tab3_reverse_2c}

In summary, there is a significant scatter in the predicted parameters
for our sample of massive galaxies by different SPS codes and IMFs. It
seems that this scatter is preferentially linked to the stellar
population codes, rather than to the IMF. BC03 (and Chabrier IMF)
provide in virtually all cases the best solutions in terms of goodness
of the fit and statistical significance of the best solution, although
the $\chi^2$ differences between models are not statistically
significant for individual galaxies. This preference is directly
linked to the SHARDS data (probing the rest-frame UV/optical range),
given that it is no longer seen when fitting broad-band fluxes only.
The stellar population in our sample of massive galaxies at z$\sim$1
is typically (according to our fiducial models, BC03-chab) 1.5~Gyr
old, with a formation time-scale around 100--200~Myr, solar or
slightly sub-solar metallicity, and moderate extinction
A(V)$\sim$0.5~mag.


\subsection{Global picture of the formation of massive galaxies at z$\sim$1}

Our sample of massive quiescent galaxies at z$>$1 is small and driven
by the availability of spectroscopic redshifts. Consequently, its
representativeness of the global population of such type of galaxies
is highly uncertain. However, it is large enough to study whether
there are significant differences in the stellar populations of
galaxies as a function of mass. This is the base for the downsizing
scenario of galaxy formation. To analyze this topic, we have divided
our sample in two mass bins with the same number of galaxies. Note
that the median mass of our sample is 10$^{10.7}$~M$_\sun$ for the
BC03-chab models, slightly smaller than the
M$^*$$\sim$10$^{10.9}$~M$_\sun$ value at z$=$1.0-1.5\footnote{Value
  obtained after accounting for differences in IMF and stellar
  population library between different works.}
\citep[][]{2008ApJ...675..234P,2009ApJ...701.1765M}, so our (limited
and not statistically representative) sample does probe the masses
around the knee of the mass function at redshift unity.

\begin{figure*}
  \begin{center}
    \hspace{-1.0cm}
    \includegraphics[angle=-90,width=9.2cm]{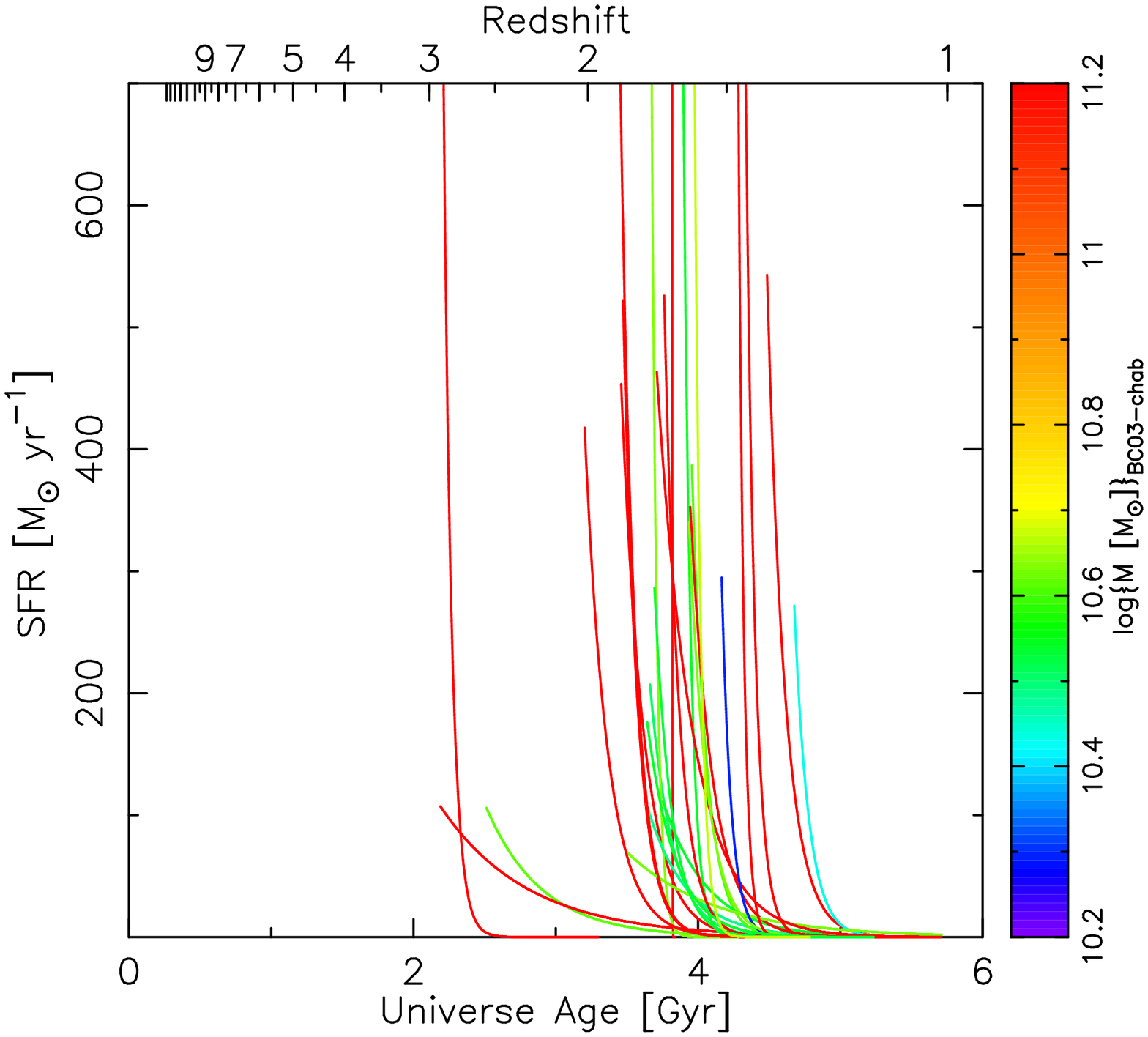}
\hspace{0.2cm}
    \includegraphics[angle=-90,width=9.2cm]{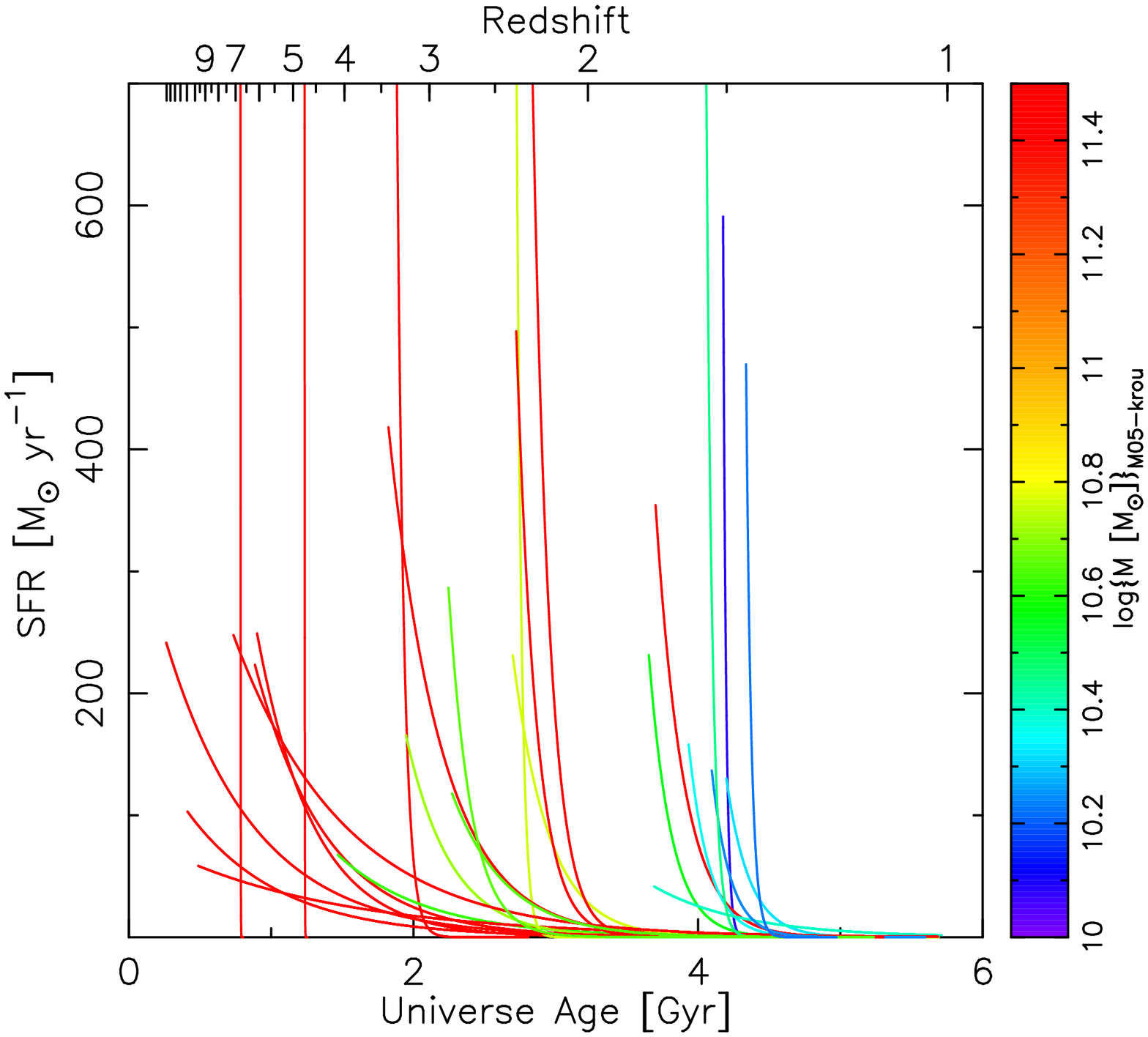}
    \figcaption{\label{fig:sfh} Star formation histories for 27
      spectroscopically confirmed quiescent massive galaxies in
      GOODS-N. The rainbow-based color code differentiates galaxies
      according to stellar mass, with the most massive sources plotted
      in yellow or redder colors, and less massive galaxies plotted in
      green or bluer colors. The left panel shows the results for the
      BC03-chab stellar population synthesis models, and the right
      plot shows the results for M05-krou.}
  \end{center}
\end{figure*}

Figure~\ref{fig:sfh} shows the SFHs of all galaxies in our sample for
the SPS results obtained with the two best-fitting models, BC03 and
M05. We have used a rainbow color code based on stellar mass: massive
galaxies are plotted in yellow and redder colors, while lighter
galaxies are depicted in green or bluer colors. According to the BC03
models, there is not a large differences in age and SFH between
galaxies of different mass. However, there is a trend for the oldest
galaxies to be among the most massive. In the case of the M05, a
segregation in mass is more evident: the lightest galaxies
preferentially present young stellar populations, while most of the
heaviest galaxies were formed 3~Gyr before the epoch of observation
(z$>$2). In addition, some of the most massive galaxies present SFHs
close to a SSP, also according to the M05 results.

The ages for the lightest half-sample range from 0.6 to 2.5~Gyr for
BC03 models, and 0.2 to 3.5~Gyr for M05. For the most massive half of
the sample, ages range from 1.0 to 2.7~Gyr according to BC03, and 2.9
to 5.0~Gyr for M05 results.  On average, the most massive half of our
sample harbors stellar populations that are 1.6~Gyr old, while the
lightest-half are 20\% younger for the results obtained with BC03
models. For M05 models, the difference is significantly larger: the
heaviest galaxies present, on average, 4~Gyr old stellar populations,
compared to 2~Gyr old stars for the lightest systems.  Other libraries
provide intermediate results, i.e., a segregation in age of galaxies
of different masses with larger differences than those for the BC03
models, but smaller than for the M05 code. If we only use broad-band
data in our fits, the results are roughly unchanged for the M05, CB09,
and P01 libraries, but the segregation is more significant for the
BC03 code: it provides an average age of 2.5~Gyr for the most massive
galaxies, and 1.5~Gyr for the lightest.

For our fiducial models, most galaxies present e-folding times around
100-300~Myr and peak SFRs around 400-500~M$_\sun$~yr$^{-1}$, but there
are some examples which are characterized by shorter bursts and could
reach even higher SFRs. For M05, e-folding times are larger
(100-600~Myr) and peak SFRs are smaller. These SFR figures and duty
cycles are typical of ULIRGs and/or SMGs at z$\sim$2
\citep[e.g.,][]{1999MNRAS.302..632B,2002MNRAS.331..495S,2005ApJ...622..772C,
  2005ApJ...630...82P,2008ApJ...677..943D,2008ApJ...689..127P}.  They
are also consistent with the SFRs predicted by cosmological
simulations involving gas-rich mergers
\citep[e.g.,][]{2010MNRAS.404.1355D,2010MNRAS.401.1613N,2010MNRAS.407.1701N}.

Our results are roughly consistent with those described in
\citet{2009ApJ...706..158F} and \citet{2012arXiv1206.2360K}. In
\citet{2009ApJ...706..158F}, they used CB09 models to fit optical low
resolution spectroscopic data of intermediate redshift massive
early-type galaxies. Based on these models, they found typical ages of
2--3~Gyr for their last redshift bin (z$>$0.9), slightly dependent on
stellar mass. No major differences were found in this paper in the
star formation timescale as a function of mass. This agrees well with
our ages around 1--2~Gyr for z$\sim$1.2 (there is 1.2~Gyr difference
in look-back time between z$=$0.9 and z$=$1.2), and the very similar
e-folding times for galaxies with different masses.
\citet{2012arXiv1206.2360K}, using CB09 models fitting broad-band
data, found similar ages for massive galaxies up to
M$\sim$10$^{10.5}$~M$_\sun$, and a trend at higher masses where
heavier galaxies are older.  However, we stress that our study points
out that the mass-age relationship should be taken with caution, given
that it is highly dependent on the SPS code used in the SED fitting.




\section{Summary and conclusions}


We have presented the basic characteristics of the Survey for High-z
Absorption Red and Dead Sources (SHARDS), an ESO/GTC Large Program
awarded 180 hours of observing time with the OSIRIS instrument on the
10.4m Gran Telescopio Canarias (GTC). The SHARDS project was devised
to be able to measure absorption indices such as the \Mg\, or D(4000)
for galaxies at z$=$0.0--2.5 through imaging data, and detect
emission-line galaxies up to z$\sim$6. For those purposes, SHARDS is
obtaining imaging data in the GOODS-N field through 24 medium-band
filters covering the wavelength range between 500 and 950~nm and
reaching magnitude 26.5 in all bands. In addition, virtually all the
images were obtained under sub-arcsec seeing conditions.

In this paper, we have presented the special reduction and calibration
procedures used to analyze the SHARDS observations. The flux
calibration has been carried out by comparison with HST and
ground-based (Keck, CAHA 3.5m telescope) spectra. The zeropoints have
also been tested by comparing with synthetic magnitudes obtained from
stellar population synthesis models fitting broad-band data. Overall,
our procedure achieves an absolute photometric calibration uncertainty
lower than 8\% (typically 5\%-7\%).

In this paper, we have carried our a science verification of the main
goals of SHARDS. We have shown how the SHARDS data are able to detect
emission lines for low, intermediate and high redshift sources.  We
have demonstrated that the depth and image quality of our survey allow
us to select virtually all emission-line galaxies which have been
already confirmed by the deepest spectroscopic surveys carried out in
the GOODS-N field, such as TKRS and DEEP3. We have shown that we are
able to extend these spectroscopic studies of emission-line galaxies
to faint magnitudes (fainter than the typical spectroscopic limit,
$RI$$\sim$24-25~mag) and detect, for example, Ly-$\alpha$ emitters at
z$\sim$5. By combining all the SHARDS data and fitting stellar
population synthesis models, we have also shown that we can measure
equivalent widths and fluxes for those emission lines.  A
comprehensive study of the population of emission-line galaxies
detected by SHARDS in GOODS-N will be presented in a future paper
(Cava et al. 2012, in preparation).

We have assessed the ability of the SHARDS data to detect, measure and
model absorption features in intermediate and high-z galaxies. For
that purpose, we have analyzed the SHARDS data for all passively
evolving galaxies at z$>$1 counting with a spectroscopic redshift,
whose SED should be dominated by absorption features. We have
constructed a sample of 27 quiescent massive galaxies with
spectroscopic redshifts at 1.01$<$z$<$1.43 extracted from several
sources and covered by the SHARDS data (and detected in all SHARDS
bands, although this was not a requirement of the sample selection).
For this sample, we have analyzed the entire spectral energy
distribution from the UV to the MIR, including broad-band data from
ground-based telescopes, HST, and {\it Spitzer}, as well as the SHARDS
photometry. This analysis consists in fitting the SEDs with stellar
population synthesis models assuming an exponentially decreasing SFH
($\tau$ models) and an extinction law described by Calzetti's law. We
have not imposed any {\it a priori} limitations on the metallicity,
$\tau$ value or age. In order to study the robustness of the derived
parameters, we have used a variety of stellar population libraries,
namely: BC03, M05, CB09, and P01. Our fitting method includes a
procedure to probe the typical degeneracies in this kind of study,
such as those linked to age-metallicity, or age-extinction. This
procedure is based on a Montecarlo method which assigns statistical
significances to the different possible solutions found by the
synthesis models.

Our results demonstrate that including the SHARDS spectro-photometric
data in the analysis of the stellar populations in high-z passively
evolving galaxies helps to break these degeneracies, although we do
not fully succeed in this task. When using the SHARDS data, the
statistical significance of the best solution (that with the highest
significance) is typically 10-20\% larger than when fitting broad-band
data alone. Typical values of the statistical significance of the best
solution when fitting all available photometry is 75\% (i.e., one
solution is favored at the 75\% probability level). Interestingly,
among the different stellar population synthesis libraries used in
this paper, BC03 models (with a Chabrier IMF) provide the best fits
(the ones with the best reduced $\chi^2$ values) for more than 90\% of
the sample when fitting SHARDS' and broad-band data. When only fitting
broad-band data, all models are (roughly) equally good in reproducing
the data. We conclude that BC03 models are the most suited to study
the rest-frame UV/optical part (up to $\sim$400~nm) of the SEDs for
our sample.

Taking into account all different models, we find that the stellar
populations in our sample of passively evolving massive sources at
z$>$1 are typically 1.5-2.0~Gyr old, presenting typical $\tau$ values
around 100--200~Myr, solar or slightly sub-solar metallicity, and
$V$-band extinctions around 0.5~mag. However, significant systematic
differences from one model library to another are found for individual
sources, especially in the age and $\tau$ values. Stellar masses are
better behaved, with the largest differences from one model to another
being below 0.3dex. Using an average of all the models, the typical
masses of the galaxies in our sample are in the interval
10$^{10.5-11.5}$~M$_\sun$.

Dividing our sample of massive quiescent galaxies at z$>$1 in half
according to their stellar masses, we find that the most massive
galaxies tend to be older for all libraries. The differences are small
for BC03 results, and considerably larger for fits using M05, CB09 or
P01 models. Just using broad-band data also produces a clearer age
segregation in mass.  Typically, for the M05 models, the lightest
galaxies present ages around 2~Gyr, while the heaviest systems are
3-4~Gyr old on average, with some galaxies reaching ages around 5 Gyr
(i.e., they started forming less than 1~Gyr after the Big Bang). No
significant difference in the SFH is found between the most massive
and lighter samples.

\section*{Acknowledgments}

We acknowledge support from the Spanish Programa Nacional de
Astronom\'{\i}a y Astrof\'{\i}sica under grants AYA2009-07723-E and
AYA2009-10368. SHARDS has been funded by the Spanish MICINN/MINECO
under the Consolider-Ingenio 2010 Program grant CSD2006-00070: First
Science with the GTC. OG-M, CM-T, JMR-E, and JR-Z wish to acknowledge
support from grant AYA2010-21887-C04-04. AA-H and AH-C acknowledge
financial support from the Universidad de Cantabria through the
Augusto G. Linares Program. This work has made use of the Rainbow
Cosmological Surveys Database, which is operated by the Universidad
Complutense de Madrid (UCM). This research has made use of the VizieR
catalogue access tool, CDS, Strasbourg, France. Based on observations
made with the Gran Telescopio Canarias (GTC), installed at the Spanish
Observatorio del Roque de los Muchachos of the Instituto de
Astrof\'{\i}sica de Canarias, in the island of La Palma. We thank all
the GTC Staff for their support and enthusiasm with the SHARDS
project, and we would like to especially acknowledge the help from
Antonio Cabrera and Ren\'e Rutten. Also based on observations
collected at the Centro Astron\'omico Hispano Alem\'an (CAHA) at Calar
Alto, operated jointly by the Max-Planck Institut f\"{u}r Astronomie
and the Instituto de Astrof\'{\i}sica de Andaluc\'{\i}a (CSIC). We
thank the referee for her/his very useful and positive comments.



\bibliographystyle{apj}
\bibliography{referencias}

\label{lastpage}
\end{document}

%% file: tab1_2c.tex
\begin{deluxetable*}{lcccccccccccccccccc}
\tabletypesize{\scriptsize}
\tablecaption{\label{table:filters}Characteristics of the SHARDS filter set and observations (before 2012A).}
\tablehead{ \colhead{Filter} & \colhead{CWL} & \colhead{Width} & \colhead{A} & \colhead{B} & \colhead{$X_0$} & \colhead{$Y_0$} & \colhead{rms} & \multicolumn{2}{c}{$\Delta$ZP}& & \multicolumn{2}{c}{m$_\mathrm{3\sigma}$} & & \multicolumn{2}{c}{m$_\mathrm{75\%}$} & & \multicolumn{2}{c}{seeing}\\
\cline{9-10}                          \cline{12-13}                                \cline{15-16}                     \cline{18-19}        \\
&                   &                 &             &             &                 &                 &               &    P1 &  P2             &      &          P1      &      P2       &        &            P1      &      P2  &        &      P1  &   P2           \\
\colhead{(1)} & \colhead{(2)} & \colhead{(3)} & \colhead{(4)} & \colhead{(5)} & \colhead{(6)} & \colhead{(7)} & \colhead{(8)} &  \colhead{(9)} &  \colhead{(9)} & &  \colhead{(10)} & \colhead{(10)} & & \colhead{(11)} & \colhead{(11)} & & \colhead{(12)} & \colhead{(12)}}
\startdata
F619W17  & 618.9 & 15.1 & 623.14 & -2.402e-6 &  -202 &  985 & 0.10 & 0.077 & \nodata & & 26.89 & \nodata & & 27.22 & \nodata & & 0.85 & \nodata\\
F636W17  & 638.4 & 15.4 & 641.37 & -2.587e-6 &  -116 &  986 & 0.10 & 0.066 & 0.072 & & 26.78 & 26.70 & & 27.15 & 27.17 & & 0.79 & 0.92\\
F653W17  & 653.1 & 14.8 & 656.01 & -2.634e-6 &  -151 &  999 & 0.10 & 0.063 & 0.065 & & 26.91 & 27.07 & & 27.17 & 27.15 & & 0.98 & 1.00\\
F670W17  & 668.4 & 15.3 & 671.86 & -2.600e-6 &  -183 & 1037 & 0.10 & 0.062 & 0.064 & & 26.64 & 26.76 & & 27.12 & 27.18 & & 0.79 & 1.07\\
F687W17  & 688.2 & 15.3 & 690.50 & -2.674e-6 &  -186 &  983 & 0.10 & 0.050 & 0.055 & & 27.04 & 26.85 & & 27.10 & 27.09 & & 0.84 & 0.93\\
F704W17  & 704.5 & 17.1 & 707.78 & -2.723e-6 &  -209 & 1028 & 0.10 & 0.055 & 0.063 & & 26.71 & 26.63 & & 27.01 & 26.95 & & 0.89 & 0.92\\
F721W17  & 720.2 & 18.2 & 723.12 & -2.969e-6 &   -94 &  960 & 0.10 & 0.056 & 0.060 & & 26.60 & 26.51 & & 26.97 & 26.96 & & 0.93 & 1.02\\
F738W17  & 737.8 & 15.0 & 741.80 & -2.411e-6 &  -328 & 1050 & 0.10 & 0.050 & 0.061 & & 26.45 & 26.25 & & 26.95 & 26.87 & & 0.86 & 0.90\\
F755W17  & 754.5 & 14.8 & 758.12 & -2.660e-6 &  -228 & 1034 & 0.10 & 0.050 & 0.055 & & 26.69 & 26.37 & & 26.93 & 26.91 & & 0.92 & 0.93\\
F772W17  & 770.9 & 15.4 & 774.62 & -2.929e-6 &  -122 & 1026 & 0.10 & 0.054 & 0.053 & & 26.54 & 26.34 & & 26.90 & 26.83 & & 0.94 & 1.04\\
F789W17  & 789.0 & 15.5 & 791.22 & -3.087e-6 &  -123 &  994 & 0.10 & 0.054 & 0.056 & & 26.22 & 26.02 & & 26.84 & 26.63 & & 0.97 & 0.92\\
F806W17  & 805.6 & 15.6 & 809.42 & -2.939e-6 &  -200 &  933 & 0.10 & 0.051 & 0.050 & & 26.39 & 26.38 & & 26.82 & 26.77 & & 0.96 & 0.99\\
F823W17  & 825.4 & 14.7 & 829.15 & -3.055e-6 &  -153 &  888 & 0.10 & 0.049 & 0.049 & & 26.59 & 26.65 & & 26.91 & 26.89 & & 0.82 & 0.91\\
F840W17  & 840.0 & 15.4 & 843.51 & -3.103e-6 &  -237 &  992 & 0.10 & 0.057 & 0.056 & & 26.13 & 26.19 & & 26.79 & 26.74 & & 0.88 & 0.91\\
F857W17  & 856.4 & 15.8 & 859.97 & -2.892e-6 &  -249 & 1002 & 0.10 & 0.050 & 0.051 & & 26.84 & 26.23 & & 26.88 & 26.63 & & 0.73 & 0.95\\
F883W35  & 880.3 & 31.7 & 885.33 & -2.889e-6 &  -285 &  978 & 0.10 & 0.065 & 0.057 & & 26.06 & 26.06 & & 26.64 & 26.58 & & 0.93 & 1.02\\
\enddata
\tablecomments{(1) Filter name. (2) Central wavelength (in nm) of the filter for angle of incidence AOI=10.5$^\circ$ (approximately that for the center of the FOV). (3) Filter width (in nm). (4) Coefficient A (in nm) for CWL calibration (from equation \ref{equ:cwl}). (5) Coefficient B (in pixel$^{-2}$) for CWL calibration (from equation \ref{equ:cwl}). (6) X coordinate for the OSIRIS optical center (in pixels). (7) Y coordinate for the OSIRIS optical center (in pixels). (8) RMS of the CWL calibration (in nm). (9) Zeropoint uncertainty for pointings 1 and 2. (10) Sensitivity limit at 3-$\sigma$ level (AB mag) for pointings 1 and 2. (11)  Third quartil of the magnitude distribution (AB mag) for pointings 1 and 2. (12) Average seeing (in arcsec) for pointings 1 and 2.}
\end{deluxetable*}

%% file: tab2_2c.tex
\placetable{table:massive}
\tabletypesize{\scriptsize}
\begin{deluxetable*}{lclrccccrr}
\tablecaption{\label{table:massive}Stellar population synthesis results for red and dead galaxies at z$\sim$1.}
\tablehead{ \colhead{Galaxy} & \colhead{z} & \colhead{Model} & \colhead{$\tau$} & \colhead{Age} & \colhead{A(V)} & \colhead{Z} & \colhead{M} & \colhead{$\chi^2$} & \colhead{Prob.}\\ 
\colhead{(1)} & \colhead{(2)} & \colhead{(3)} & \colhead{(4)} & \colhead{(5)} & \colhead{(6)} & \colhead{(7)} & \colhead{(8)} & \colhead{(9)} & \colhead{(10)}}
\startdata
J123704.36+621335.03 & 1.2420 &  BC03-chab  & $158_{-16}^{+18}$ & $1.43_{-0.13}^{+0.16}$ & $0.40_{-0.04}^{+0.05}$ & $1.000_{-0.601}^{+0.000}$ & $10.86_{-0.04}^{+0.05}$ & $0.16_{-0.14}^{+0.18}$ & 76.1\\
                     &        &  CB09-chab  & $   32_{-5}^{+5}$ & $1.02_{-0.11}^{+0.10}$ & $0.30_{-0.04}^{+0.05}$ & $1.000_{-0.000}^{+0.000}$ & $10.69_{-0.04}^{+0.05}$ & $0.38_{-0.35}^{+0.40}$ & 85.6\\
                     &        &  CB09-krou  & $   40_{-5}^{+5}$ & $1.02_{-0.09}^{+0.10}$ & $0.40_{-0.05}^{+0.05}$ & $1.000_{-0.000}^{+0.000}$ & $10.78_{-0.04}^{+0.04}$ & $0.37_{-0.34}^{+0.39}$ & 68.1\\
                     &        &  M05-krou   & $501_{-54}^{+51}$ & $4.00_{-0.43}^{+0.40}$ & $0.20_{-0.04}^{+0.05}$ & $0.500_{-0.000}^{+0.000}$ & $11.05_{-0.05}^{+0.05}$ & $0.22_{-0.21}^{+0.25}$ & 91.3\\
                     &        &  P01-krou   & $   50_{-6}^{+5}$ & $3.50_{-0.35}^{+0.33}$ & $0.50_{-0.03}^{+0.05}$ & $0.200_{-0.000}^{+0.000}$ & $11.16_{-0.04}^{+0.05}$ & $0.28_{-0.26}^{+0.30}$ & 56.5\\
                     &        & \multicolumn{8}{c}{}\\
                     &        &  CB09-chab  & $584_{-170}^{+94}$ & $4.43_{-0.18}^{+0.51}$ & $0.29_{-0.04}^{+0.05}$ & $0.200_{-0.000}^{+0.000}$ & $11.06_{-0.05}^{+0.05}$ & $0.38_{-0.36}^{+0.40}$ & 14.4\\
                     &        &  CB09-krou  & $494_{-58}^{+55}$ & $4.72_{-0.46}^{+0.45}$ & $0.31_{-0.06}^{+0.04}$ & $0.200_{-0.000}^{+0.000}$ & $11.16_{-0.05}^{+0.04}$ & $0.36_{-0.34}^{+0.39}$ & 31.9\\
                     &        &  P01-krou   & $347_{-45}^{+135}$ & $3.77_{-0.51}^{+0.61}$ & $0.51_{-0.06}^{+0.05}$ & $0.200_{-0.000}^{+0.000}$ & $11.13_{-0.06}^{+0.05}$ & $0.29_{-0.27}^{+0.31}$ & 18.2\\
                     &        &  P01-krou   & $227_{-42}^{+31}$ & $3.07_{-0.40}^{+0.56}$ & $0.51_{-0.05}^{+0.05}$ & $0.200_{-0.000}^{+0.000}$ & $11.10_{-0.06}^{+0.06}$ & $0.29_{-0.26}^{+0.30}$ & 16.9\\
                     &        & \multicolumn{8}{c}{}\\
                     &        & \multicolumn{8}{c}{}\\
J123615.24+620944.52 & 1.2674 &  BC03-chab  & $1.0_{-0.1}^{+0.1}$ & $1.02_{-0.10}^{+0.10}$ & $1.00_{-0.05}^{+0.06}$ & $0.200_{-0.186}^{+0.000}$ & $10.95_{-0.05}^{+0.05}$ & $0.68_{-0.46}^{+0.97}$ & 88.9\\
                     &        &  CB09-chab  & $   63_{-9}^{+10}$ & $1.02_{-0.10}^{+0.10}$ & $0.20_{-0.05}^{+0.05}$ & $1.000_{-0.000}^{+0.000}$ & $10.68_{-0.05}^{+0.05}$ & $0.83_{-0.54}^{+1.16}$ & 65.2\\
                     &        &  CB09-krou  & $126_{-15}^{+16}$ & $1.02_{-0.11}^{+0.13}$ & $0.40_{-0.05}^{+0.05}$ & $1.000_{-0.984}^{+0.000}$ & $10.77_{-0.05}^{+0.05}$ & $0.82_{-0.53}^{+1.18}$ & 67.2\\
                     &        &  M05-krou   & $   32_{-6}^{+9}$ & $3.00_{-0.27}^{+0.38}$ & $0.80_{-0.05}^{+0.05}$ & $0.020_{-0.000}^{+0.000}$ & $11.11_{-0.06}^{+0.04}$ & $0.89_{-0.61}^{+1.18}$ & 30.2\\
                     &        &  P01-krou   & $2.0_{-0.2}^{+0.3}$ & $4.50_{-0.48}^{+0.45}$ & $1.10_{-0.07}^{+0.05}$ & $0.005_{-0.000}^{+0.015}$ & $11.34_{-0.05}^{+0.05}$ & $0.84_{-0.58}^{+1.16}$ & 40.9\\
                     &        & \multicolumn{8}{c}{}\\
                     &        &  CB09-chab  & $1.0_{-0.1}^{+0.1}$ & $0.72_{-0.07}^{+0.08}$ & $0.40_{-0.05}^{+0.04}$ & $1.000_{-0.000}^{+0.000}$ & $10.67_{-0.05}^{+0.04}$ & $0.85_{-0.57}^{+1.22}$ & 34.8\\
                     &        &  CB09-krou  & $   63_{-7}^{+8}$ & $1.01_{-0.10}^{+0.12}$ & $0.31_{-0.04}^{+0.05}$ & $1.000_{-0.000}^{+1.851}$ & $10.78_{-0.04}^{+0.05}$ & $0.83_{-0.53}^{+1.20}$ & 32.8\\
                     &        &  M05-krou   & $1.0_{-0.1}^{+0.1}$ & $4.00_{-0.35}^{+0.50}$ & $0.70_{-0.05}^{+0.06}$ & $0.020_{-0.000}^{+0.000}$ & $11.17_{-0.04}^{+0.05}$ & $0.64_{-0.42}^{+0.91}$ & 29.5\\
                     &        &  M05-krou   & $   31_{-4}^{+5}$ & $2.94_{-0.34}^{+0.33}$ & $0.81_{-0.06}^{+0.05}$ & $0.020_{-0.000}^{+0.000}$ & $11.11_{-0.03}^{+0.05}$ & $0.78_{-0.52}^{+1.10}$ & 29.2\\
                     &        &  M05-krou   & $   18_{-4}^{+4}$ & $2.98_{-0.26}^{+0.62}$ & $0.80_{-0.07}^{+0.07}$ & $0.020_{-0.000}^{+0.000}$ & $11.11_{-0.04}^{+0.06}$ & $0.67_{-0.46}^{+0.97}$ & 11.1\\
                     &        &  P01-krou   & $1.0_{-0.1}^{+0.1}$ & $4.53_{-0.51}^{+0.44}$ & $0.90_{-0.04}^{+0.06}$ & $0.020_{-0.015}^{+0.000}$ & $11.30_{-0.04}^{+0.04}$ & $0.74_{-0.51}^{+1.09}$ & 37.2\\
                     &        &  P01-krou   & $   20_{-2}^{+2}$ & $4.44_{-0.39}^{+0.48}$ & $1.10_{-0.05}^{+0.03}$ & $0.005_{-0.000}^{+0.000}$ & $11.34_{-0.04}^{+0.05}$ & $0.77_{-0.55}^{+0.98}$ & 11.7\\
                     &        & \multicolumn{8}{c}{}\\
                     &        & \multicolumn{8}{c}{}\\
J123627.17+620756.57 & 1.3950 &  BC03-chab  & $398_{-43}^{+54}$ & $2.00_{-0.21}^{+0.25}$ & $0.20_{-0.05}^{+0.05}$ & $0.200_{-0.000}^{+0.000}$ & $10.62_{-0.06}^{+0.04}$ & $1.45_{-1.04}^{+1.97}$ & 56.1\\
                     &        &  CB09-chab  & $   63_{-6}^{+8}$ & $0.51_{-0.05}^{+0.07}$ & $0.20_{-0.05}^{+0.05}$ & $1.000_{-0.000}^{+0.000}$ & $10.31_{-0.05}^{+0.05}$ & $2.28_{-1.70}^{+2.95}$ & 52.5\\
                     &        &  CB09-krou  & $501_{-48}^{+65}$ & $2.60_{-0.29}^{+0.29}$ & $0.00_{-0.00}^{+0.05}$ & $0.200_{-0.180}^{+0.000}$ & $10.67_{-0.05}^{+0.05}$ & $2.28_{-1.65}^{+2.91}$ & 71.7\\
                     &        &  M05-krou   & $631_{-64}^{+80}$ & $3.00_{-0.33}^{+0.34}$ & $0.20_{-0.05}^{+0.05}$ & $0.020_{-0.000}^{+0.000}$ & $10.62_{-0.05}^{+0.05}$ & $1.68_{-1.21}^{+2.16}$ & 49.1\\
                     &        &  P01-krou   & $251_{-28}^{+37}$ & $0.99_{-0.08}^{+0.09}$ & $0.40_{-0.05}^{+0.06}$ & $1.000_{-0.603}^{+1.539}$ & $10.71_{-0.04}^{+0.05}$ & $2.23_{-1.63}^{+2.79}$ & 36.3\\
                     &        & \multicolumn{8}{c}{}\\
                     &        &  BC03-chab  & $199_{-19}^{+20}$ & $1.03_{-0.13}^{+0.13}$ & $0.46_{-0.14}^{+0.06}$ & $0.400_{-0.206}^{+0.598}$ & $10.58_{-0.04}^{+0.04}$ & $1.67_{-1.19}^{+2.25}$ & 39.1\\
                     &        &  CB09-chab  & $101_{-9}^{+10}$ & $0.73_{-0.07}^{+0.08}$ & $0.00_{-0.00}^{+0.05}$ & $1.000_{-0.615}^{+0.000}$ & $10.35_{-0.05}^{+0.04}$ & $2.17_{-1.64}^{+2.76}$ & 28.6\\
                     &        &  CB09-chab  & $   22_{-3}^{+3}$ & $0.50_{-0.06}^{+0.05}$ & $0.30_{-0.04}^{+0.03}$ & $0.400_{-0.396}^{+0.000}$ & $10.32_{-0.03}^{+0.04}$ & $1.90_{-1.42}^{+2.63}$ & 15.0\\
                     &        &  CB09-krou  & $181_{-64}^{+29}$ & $0.93_{-0.18}^{+0.15}$ & $0.14_{-0.09}^{+0.14}$ & $1.000_{-0.802}^{+0.000}$ & $10.46_{-0.05}^{+0.05}$ & $1.95_{-1.38}^{+2.64}$ & 19.8\\
                     &        &  M05-krou   & $162_{-23}^{+25}$ & $1.50_{-0.15}^{+0.20}$ & $0.00_{-0.00}^{+0.05}$ & $0.020_{-0.000}^{+3.084}$ & $10.41_{-0.05}^{+0.06}$ & $2.16_{-1.66}^{+2.85}$ & 35.4\\
                     &        &  M05-krou   & $997_{-116}^{+133}$ & $4.01_{-0.40}^{+0.49}$ & $0.29_{-0.04}^{+0.05}$ & $0.020_{-0.000}^{+0.000}$ & $10.73_{-0.05}^{+0.04}$ & $2.04_{-1.64}^{+2.63}$ & 15.5\\
                     &        &  P01-krou   & $663_{-74}^{+118}$ & $2.51_{-0.25}^{+0.23}$ & $0.22_{-0.05}^{+0.40}$ & $0.200_{-0.180}^{+0.199}$ & $10.75_{-0.05}^{+0.05}$ & $2.05_{-1.55}^{+2.69}$ & 28.6\\
                     &        &  P01-krou   & $496_{-51}^{+50}$ & $1.95_{-0.24}^{+0.27}$ & $0.12_{-0.06}^{+0.06}$ & $0.400_{-0.200}^{+0.000}$ & $10.69_{-0.04}^{+0.06}$ & $1.84_{-1.35}^{+2.50}$ & 24.7\\
                     &        &  P01-krou   & $360_{-59}^{+47}$ & $1.46_{-0.28}^{+0.19}$ & $0.32_{-0.13}^{+0.21}$ & $0.400_{-0.380}^{+0.602}$ & $10.69_{-0.06}^{+0.05}$ & $1.60_{-1.25}^{+2.12}$ & 10.4\\
                     &        & \multicolumn{8}{c}{}\\
                     &        & \multicolumn{8}{c}{}\\
J123738.71+621727.86 & 1.2907 &  BC03-chab  & $   57_{-9}^{+9}$ & $2.62_{-0.29}^{+0.29}$ & $0.20_{-0.04}^{+0.05}$ & $0.200_{-0.000}^{+0.000}$ & $11.18_{-0.04}^{+0.05}$ & $0.26_{-0.24}^{+0.27}$ & 91.1\\
                     &        &  CB09-chab  & $   63_{-6}^{+7}$ & $1.01_{-0.10}^{+0.12}$ & $0.00_{-0.00}^{+0.04}$ & $1.000_{-0.000}^{+0.000}$ & $10.83_{-0.05}^{+0.05}$ & $1.20_{-0.82}^{+1.69}$ & 100.0\\
                     &        &  CB09-krou  & $   70_{-10}^{+13}$ & $1.01_{-0.12}^{+0.10}$ & $0.10_{-0.05}^{+0.04}$ & $1.000_{-0.000}^{+0.000}$ & $10.91_{-0.04}^{+0.04}$ & $1.27_{-0.86}^{+1.79}$ & 41.6\\
                     &        &  M05-krou   & $1.0_{-0.1}^{+0.1}$ & $3.98_{-0.38}^{+0.42}$ & $0.51_{-0.05}^{+0.04}$ & $0.020_{-0.000}^{+0.000}$ & $11.31_{-0.04}^{+0.04}$ & $0.46_{-0.44}^{+0.48}$ & 100.0\\
                     &        &  P01-krou   & $500_{-60}^{+55}$ & $3.47_{-0.32}^{+0.34}$ & $0.20_{-0.04}^{+0.05}$ & $0.200_{-0.000}^{+0.000}$ & $11.23_{-0.05}^{+0.05}$ & $1.12_{-0.74}^{+1.56}$ & 72.6\\
                     &        & \multicolumn{8}{c}{}\\
                     &        &  CB09-krou  & $135_{-16}^{+20}$ & $2.65_{-0.29}^{+0.28}$ & $0.10_{-0.04}^{+0.04}$ & $0.200_{-0.000}^{+0.000}$ & $11.15_{-0.05}^{+0.05}$ & $1.15_{-0.79}^{+1.61}$ & 36.6\\
                     &        &  CB09-krou  & $1.0_{-0.1}^{+0.1}$ & $1.03_{-0.13}^{+0.09}$ & $0.00_{-0.00}^{+0.03}$ & $1.000_{-0.000}^{+0.000}$ & $10.88_{-0.04}^{+0.04}$ & $0.97_{-0.69}^{+1.31}$ & 16.0\\
                     &        &  P01-krou   & $257_{-30}^{+39}$ & $2.55_{-0.26}^{+0.43}$ & $0.29_{-0.05}^{+0.06}$ & $0.200_{-0.181}^{+0.000}$ & $11.21_{-0.04}^{+0.05}$ & $1.24_{-0.85}^{+1.75}$ & 15.9\\
                     &        &  P01-krou   & $1.0_{-0.1}^{+0.1}$ & $3.00_{-0.23}^{+0.43}$ & $0.10_{-0.05}^{+0.05}$ & $0.200_{-0.000}^{+0.000}$ & $11.21_{-0.05}^{+0.06}$ & $0.96_{-0.68}^{+1.61}$ & 11.5\\
                     &        & \multicolumn{8}{c}{}\\
                     &        & \multicolumn{8}{c}{}\\
\enddata
\tablecomments{Results of the stellar population synthesis modeling
  of the SEDs of massive quiescent galaxies at 1.0$<$z$<$1.4. Averages
  and $\pm$1$\sigma$ values are quoted for each parameter. First five rows
  for each galaxy present the results for the most statistically
  significant solution for each one of the SPS libraries presented in the
  text. Next, secondary solutions are given (if available). 
  (1) Name of the galaxy (including J2000 coordinates). 
  (2) Spectroscopic redshift from the literature
  (see text for references). (3) Stellar population models and IMF.
  (4) Exponential decay factor and
  uncertainty of the best fitting model (in Myr). (5) Age and
  uncertainty (in Gyr) of the best fitting model. (6) Extinction in
  the $V$-band and uncertainty (in mag) of the best fitting model.
  (7) Metallicity and uncertainty (solar units) of the best fitting model.
  (8) Stellar mass and uncertainty (solar units).
  (9) Median and 68\% range for the goodness of the fit ($\chi^2$ value).
  (10) Statistical significance of this solution (in \%); only 
  solutions with a significance above 10\% are shown. 
  This table includes all galaxies in the electronic version of the paper.}
\end{deluxetable*}

%% file: tab3_reverse_2c.tex
\placetable{table:stats}
\tablewidth{400pt}
\begin{deluxetable*}{lcrrrrr}
\tablecaption{\label{table:stats}Statistical properties of the stellar populations in z$\sim$1 massive quiescent galaxies.}
\tablehead{\colhead{Parameter} & \colhead{sample} & \colhead{BC03-chab} & \colhead{CB09-chab} & \colhead{CB09-krou} & \colhead{M05-krou} & \colhead{P01-krou}}
\startdata
M [M$_\sun$] &  all  & $10.67_{10.53}^{10.96}$ & $10.50_{10.35}^{10.82}$ & $10.67_{10.47}^{10.91}$ & $10.77_{10.36}^{11.19}$ & $10.79_{10.65}^{11.18}$ \\
             & light & $10.54_{10.49}^{10.63}$ & $10.38_{10.32}^{10.49}$ & $10.53_{10.43}^{10.62}$ & $10.51_{10.25}^{10.72}$ & $10.70_{10.57}^{10.75}$ \\
             & heavy & $10.86_{10.74}^{11.11}$ & $10.69_{10.66}^{10.89}$ & $10.79_{10.74}^{10.98}$ & $11.11_{10.84}^{11.32}$ & $11.15_{10.88}^{11.33}$ \\
$\tau$ [Myr] &  all  & $119_{50}^{280}$ & $ 63_{ 9}^{151}$ & $110_{38}^{436}$ & $162_{36}^{632}$ & $191_{23}^{500}$ \\
             & light & $123_{50}^{309}$ & $ 62_{40}^{157}$ & $ 88_{15}^{494}$ & $158_{36}^{380}$ & $185_{25}^{396}$ \\
             & heavy & $103_{54}^{173}$ & $ 65_{ 7}^{129}$ & $122_{61}^{194}$ & $375_{42}^{663}$ & $193_{19}^{909}$ \\
Age [Gyr]    &  all  & $1.4_{1.0}^{2.0}$ & $1.0_{1.0}^{2.2}$ & $1.0_{1.0}^{2.6}$ & $3.0_{1.5}^{4.0}$ & $1.6_{1.2}^{3.5}$ \\
             & light & $1.4_{1.0}^{1.9}$ & $1.1_{0.5}^{1.4}$ & $1.6_{1.0}^{2.6}$ & $2.0_{0.9}^{3.0}$ & $1.5_{1.2}^{1.8}$ \\
             & heavy & $1.6_{1.0}^{2.3}$ & $1.9_{1.0}^{3.2}$ & $1.5_{1.0}^{1.9}$ & $3.6_{3.0}^{4.1}$ & $2.6_{1.4}^{3.8}$ \\
A(V) [mag]   &  all  & $0.50_{0.23}^{0.94}$ & $0.30_{0.10}^{0.50}$ & $0.40_{0.01}^{0.70}$ & $0.40_{0.20}^{0.80}$ & $0.50_{0.35}^{0.81}$ \\
             & light & $0.48_{0.24}^{0.93}$ & $0.45_{0.10}^{0.50}$ & $0.27_{0.00}^{0.59}$ & $0.28_{0.20}^{0.40}$ & $0.50_{0.36}^{0.69}$ \\
             & heavy & $0.50_{0.22}^{0.92}$ & $0.30_{0.09}^{0.51}$ & $0.40_{0.10}^{0.72}$ & $0.51_{0.37}^{0.84}$ & $0.50_{0.28}^{1.09}$ \\
Z [Z$_\sun$] &  all  & $1.000_{0.200}^{2.500}$ & $1.000_{0.400}^{1.000}$ & $1.000_{0.400}^{2.500}$ & $0.500_{0.020}^{1.000}$ & $1.000_{0.200}^{2.500}$ \\
             & light & $1.000_{0.200}^{2.500}$ & $1.000_{0.400}^{1.000}$ & $1.000_{0.400}^{2.500}$ & $0.750_{0.500}^{1.000}$ & $1.000_{1.000}^{2.458}$ \\
             & heavy & $1.000_{0.200}^{2.500}$ & $1.000_{0.400}^{1.192}$ & $1.000_{0.374}^{2.500}$ & $0.020_{0.020}^{1.000}$ & $0.400_{0.020}^{2.500}$ \\
\enddata
\tablecomments{Median values and quartiles for the distribution
  of stellar parameters of the sample of 27
  massive quiescent galaxies at z$=$1.0--1.4 obtained by fitting different
  stellar population models to broad-band and SHARDS data. Statistics
  are given for all the sample and for the two half-samples obtained by breaking it
  by stellar mass (using the median value for the results obtained with each model).}
\end{deluxetable*}